%% file: main.tex
\documentclass[10pt,journal,compsoc]{IEEEtran}

\usepackage[normalem]{ulem}

\input{macros}

\begin{document}

\title{\huge PANTHER: A Programmable Architecture for Neural Network Training Harnessing Energy-efficient ReRAM}



\author{
        Aayush Ankit, 
        Izzat El Hajj, 
        Sai Rahul Chalamalasetti, 
        Sapan Agarwal, Matthew Marinella,
        Martin Foltin, John Paul Strachan, Dejan Milojicic, 
        Wen-mei Hwu, 
        and Kaushik Roy.%
\IEEEcompsocitemizethanks{ 
    \IEEEcompsocthanksitem A. Ankit, and K. Roy are with the Department of Electrical and Computer Engineering, Purdue University.
    \IEEEcompsocthanksitem I. E. Hajj is with the Department of Computer Science, American University of Beirut.
    \IEEEcompsocthanksitem S. R. Chalamalasetti, M. Foltin, J. P. Strachan, and D. Milojicic are with the Hewlett Packard Labs.
    \IEEEcompsocthanksitem S. Agarwal, and M. Marinella are with the Sandia National Labs.
    \IEEEcompsocthanksitem W. Hwu is with the Department of Electrical and Computer Engineering, University of Illinois at Urbana-Champaign.
    }
\thanks{Correspondance email: aankit@purdue.edu}
}

\IEEEtitleabstractindextext{
\input{sec/00-abstract.tex}
}

\maketitle

\thispagestyle{empty}

\input{sec/01-introduction.tex}

\input{sec/02-background.tex}

\input{sec/03-op-precision.tex}

\input{sec/04-mcu.tex}

\input{sec/05-accelerator.tex}

\input{sec/06-methodology.tex}

\input{sec/07-evaluation.tex}

\input{sec/08-related.tex}

\input{sec/09-conclusion.tex}

\input{sec/10-ack.tex}

\vspace{-6mm}
\balance
\bibliographystyle{unsrt}
\bibliography{ref}

\end{document}

%% file: macros.tex
\usepackage{balance}

\newcommand{\hatm}[1]{\expandafter\hat#1}

\usepackage{mathtools}
\usepackage{multirow}

\usepackage{pifont}

\usepackage{xspace}
\newcommand{\MTVM}{M\textsuperscript{T}VM\xspace}

\newcommand{\cmosall}{\textit{Base\textsubscript{digital}}}
\newcommand{\rerammvm}{\textit{Base\textsubscript{mvm}}}
\newcommand{\reramopasmvm}{\textit{Base\textsubscript{opa/mvm}}}
\newcommand{\reramop}{PANTHER}

\newcommand{\shrinkBeforeCaption}{\vspace{-6mm}}
\newcommand{\shrinkAfterCaption}{\vspace{-4mm}}

\newcommand{\ignore}[1]{}

\usepackage[dvipsnames]{xcolor}

\newif\ifsubmit
\submittrue
\ifsubmit
    \newcommand{\aayush}[1]{}
    \newcommand{\izzat}[1]{}
    \newcommand{\sai}[1]{}
    \newcommand{\sapan}[1]{}
    \newcommand{\matt}[1]{}
    \newcommand{\martin}[1]{}
    \newcommand{\johnpaul}[1]{}
    \newcommand{\dejan}[1]{}
    \newcommand{\wenmei}[1]{}
    \newcommand{\kaushik}[1]{}
    \newcommand{\todo}[1]{}
    \newcommand{\tocite}[1]{}
    \newcommand{\addition}[1]{#1}
    \newcommand{\additionx}[1]{#1}
\else
    \newcommand{\aayush}[1]{[{\color{cyan}AA: #1}]}
    \newcommand{\izzat}[1]{[{\color{green}IE: #1}]}
    \newcommand{\sai}[1]{[{\color{purple}SC: #1}]}
    \newcommand{\sapan}[1]{[{\color{purple}SA: #1}]}
    \newcommand{\matt}[1]{[{\color{purple}MM: #1}]}
    \newcommand{\martin}[1]{[{\color{purple}MF: #1}]}
    \newcommand{\johnpaul}[1]{[{\color{purple}JPS: #1}]}
    \newcommand{\dejan}[1]{[{\color{violet}DM: #1}]}
    \newcommand{\wenmei}[1]{[{\color{orange}WMH: #1}]}
    \newcommand{\kaushik}[1]{[{\color{purple}KR: #1}]}
    \newcommand{\todo}[1]{[{\color{red}TODO: #1}]}
    \newcommand{\tocite}[1]{[{\color{red}CITE: #1}]}
    \newcommand{\addition}[1]{{\color{brown} #1}}
    \newcommand{\additionx}[1]{{\color{blue} #1}}
\fi

%% file: sec/00-abstract.tex
\begin{abstract}

The wide adoption of deep neural networks has been accompanied by ever-increasing energy and performance demands due to the expensive nature of training them.
Numerous special-purpose architectures have been proposed to accelerate training: both digital and hybrid digital-analog using resistive RAM (ReRAM) crossbars.
ReRAM-based accelerators have demonstrated the effectiveness of ReRAM crossbars at performing \additionx{matrix-vector multiplication} operations that are prevalent in training.
However, they still suffer from inefficiency due to the use of serial reads and writes for performing the weight gradient and update step.

A few works have demonstrated the possibility of performing outer products in crossbars, which can be used to realize the weight gradient and update step without the use of serial reads and writes.
However, these works have been limited to low precision operations which are not sufficient for typical training workloads.
Moreover, they have been confined to a limited set of training algorithms for fully-connected layers only.

To address these limitations, we propose a bit-slicing technique for enhancing the precision of ReRAM-based outer products, \additionx{which is substantially different from bit-slicing for matrix-vector multiplication only}.
We incorporate this technique into a crossbar architecture with three variants catered to different training algorithms.
\additionx{To evaluate our design on different types of layers in neural networks (fully-connected, convolutional, etc.) and training algorithms, we develop \reramop{}, an ISA-programmable training accelerator with compiler support.}
Our design can also be integrated into other accelerators in the literature to enhance their efficiency.
Our evaluation shows that \additionx{\reramop{}} achieves up to $8.02\times$, $54.21\times$, and $103\times$ energy reductions as well as $7.16\times$, $4.02\times$, and  $16\times$ execution time reductions compared to digital accelerators, ReRAM-based accelerators, and GPUs, respectively.

\end{abstract}

%% file: sec/01-introduction.tex
\section{Introduction}

Deep Neural Networks (DNNs) have seen wide adoption due to their success in many domains such as image processing, speech recognition, and natural language processing.
However, DNN training requires substantial amount of computation and energy which has led to the emergence of numerous special-purpose accelerators~\cite{sze2017efficient}.
These accelerators have been built using various circuit technologies, including digital CMOS logic~\cite{chen2014dadiannao, jouppi2017tpu} as well as hybrid digital-analog logic based on ReRAM crossbars~\cite{shafiee2016isaac, chi2016prime}.

ReRAM crossbars are circuits composed of non-volatile elements that can perform Matrix-Vector Multiplication (MVM) in the analog domain with low latency and energy consumption.
Since MVM operations dominate the performance of DNN inference and training, various inference~\cite{shafiee2016isaac,chi2016prime,ankit2019puma} and training~\cite{cheng2017time,song2017pipelayer} accelerators have been built using these crossbars.
However, while inference algorithms do not modify matrices during execution, training algorithms modify them during the weight gradient and update step (weight gradient computation followed by the weight update).
For this reason, training accelerators~\cite{cheng2017time,song2017pipelayer} require frequent reads and write to crossbar cells to realize weight gradient and update operations.
These reads and writes to ReRAM crossbars are performed one row at a time (like a typical memory array), and are referred to as \textit{serial reads and writes} in this paper.

\begin{figure}[t]
  \centering
  \includegraphics[width=\columnwidth]{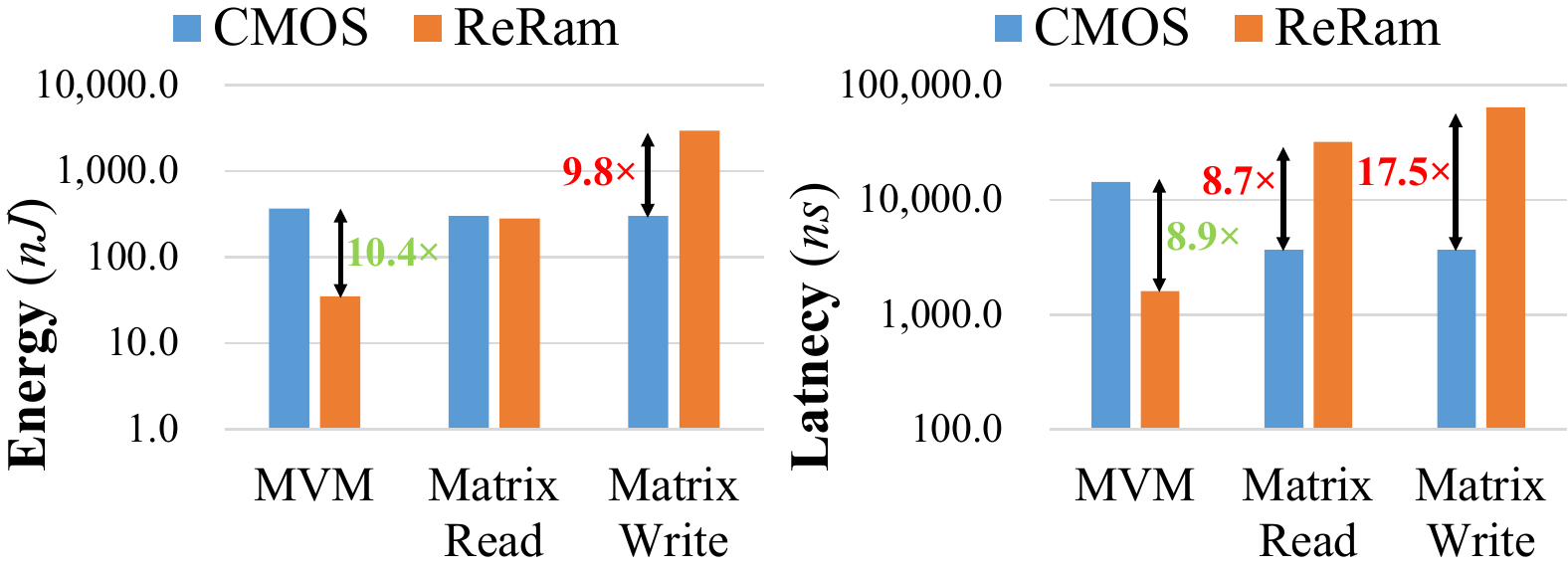}
  \shrinkBeforeCaption
  \caption{Comparing CMOS and ReRAM Primitives}\label{fig:write-overhead}
  \shrinkAfterCaption
\end{figure}

Figure~\ref{fig:write-overhead} compares the energy and latency of CMOS and ReRAM technologies for various primitive operations.
As shown, MVM consumes $\simeq10.4\times$ less energy and has $\simeq8.9\times$ lower latency with ReRAM over CMOS (at same area) for a 32 nm technology node.
However, reading and writing the entire matrix consumes much higher energy and latency with ReRAM.
Particularly, ReRAM writing energy and latency are an order of magnitude higher due to the cost of the program-verify approach which requires tens of pulses~\cite{merced2016repeatable}.
Therefore, the use of serial reads and writes during training takes away the overall benefits gained from using ReRAM for acceleration.

To overcome this issue, recent demonstrations~\cite{marinella2018multiscale,narayanan2017toward} have shown that Outer Product Accumulate (OPA) operations can be performed in crossbars to realize the weight gradient and update operations without the use of serial reads and writes.
The OPA operation is performed by applying two input vectors at the rows and the columns of a crossbar simultaneously, to update each cell depending on the inputs at the corresponding row and column. 
However, these demonstrations are limited to low-precision inputs/outputs (2-4 bits) and weights (2-5 bits) which is not sufficient for the typical training workloads~\cite{micikevicius2017mixed,wu2018training}.
Moreover, they are confined to Stochastic Gradient Descent (SGD) with batch size of one for fully-connected layers only.

To address these limitations, we propose a bit-slicing technique for achieving higher precision OPA operations by slicing the bits of the output matrix weights across multiple crossbars.
While bit-slicing has previously been done for MVM operations~\cite{shafiee2016isaac}, bit-slicing matrices to also support OPA operations is substantially different.
For MVM, the rows and the crossbar cells are inputs and the columns are outputs, whereas for OPA, the rows and the columns are both inputs and the outputs are the crossbar cells themselves.
Moreover, bit-slicing OPA presents additional constraints for the distribution of bits across the slices.
First, weights are constant during MVM, but they change during OPA, which necessitates support for overflow within each slice \additionx{and accounting for saturation}.
Second, MVM favors fewer bits per slice to reduce analog-to-digital Converter (ADC) precision requirements~\cite{shafiee2016isaac}, but we show that OPA favors more bits per slice.
Third, MVM favors homogeneous slicing of bits (equal number of bits per slice), but we show that OPA favors heterogeneous slicing.

We incorporate our proposed technique for enhancing OPA precision into a crossbar architecture that performs both MVM and OPA operations at high precision.
We present three variants of the crossbar architecture that are catered to different training algorithms: SGD, mini-batch SGD, and mini-batch SGD with large batches.
\additionx{
Using this crossbar architecture, we build \reramop{}, a \underline{P}rogrammable \underline{A}rchitecture for \underline{N}eural Network \underline{T}raining \underline{H}arnessing \underline{E}nergy-efficient \underline{R}eRAM.
We use \reramop{}
}
to evaluate our design on different layer types (fully-connected, convolutional, etc.) and training algorithms.
\additionx{Our design can also be integrated into existing training accelerators in the literature to enhance their efficiency.}
Our evaluation shows that \additionx{\reramop{}} achieves up to $8.02\times$, $54.21\times$, and $2,358\times$ energy reductions as well as $7.16\times$, $4.02\times$, and  $119\times$ execution time reductions compared to digital accelerators, ReRAM-based accelerators, and GPUs, respectively.

We make the following contributions:
\vspace{-2mm}
\begin{itemize}\setlength\itemsep{-0.1em}
    \item \additionx{A} bit-slicing technique for implementing high-precision OPA operations using ReRAM crossbars (Section~\ref{sec:op-precision})
    \item \additionx{A} crossbar-based architecture, that embodies this bit-slicing technique, with three variants for different training algorithms (Section~\ref{sec:mcu})
    \item \additionx{An} ISA-programmable accelerator with compiler support to evaluate different types of layers in neural networks and training algorithms (Section~\ref{sec:accelerator})
\end{itemize}
\vspace{-2mm}
We begin with a background on the use of ReRAM crossbars for DNN training (Section~\ref{sec:background}).

%% file: sec/02-background.tex
\section{Background}\label{sec:background}

\subsection{Deep Neural Network Training}

Typical DNN training comprises of iterative updates to a model's weights in order to optimize the loss based on an objective function.
Equations~\ref{eqn:background-dnn-fw}--\ref{eqn:background-dnn-bw-weight-grad} show the steps involved in DNN training based on the Stochastic Gradient Descent (SGD) algorithm~\cite{rumelhart1985learning}.
Equation~\ref{eqn:background-dnn-fw} constitutes the forward pass which processes an input example to compute the activations at each layer.
Equation~\ref{eqn:background-dnn-loss} computes the output error and its gradient based on a loss function using the activations of the final layer.
Equations~\ref{eqn:background-dnn-bw-layer-grad} constitutes the backward pass which propagates the output error to compute the errors at each layer.
Finally, equation~\ref{eqn:background-dnn-bw-weight-grad} computes the weight updates to minimize the error.

\vspace{-2mm}
\begin{equation} \label{eqn:background-dnn-fw}
    	H\textsuperscript{(l+1)} = W\textsuperscript{(l)} \ X\textsuperscript{(l)}, \
    	X\textsuperscript{(l+1)} = \sigma(H\textsuperscript{(l+1)}) \\
\end{equation}
\vspace{-9mm}

\begin{equation} \label{eqn:background-dnn-loss}
    	E = Loss(X\textsuperscript{(L)}, y), \
    	\delta H\textsuperscript{(L)} = \nabla E \odot \sigma\textsuperscript{$\prime$}(X\textsuperscript{(L)}) \\
\end{equation}
\vspace{-9mm}

\begin{equation} \label{eqn:background-dnn-bw-layer-grad}
    	\delta H\textsuperscript{(l)} = [(W\textsuperscript{l})\textsuperscript{T} \  \delta H\textsuperscript{(l+1)}]  \odot  \sigma\textsuperscript{$\prime$}(X\textsuperscript{(l)}) \\
\end{equation}
\vspace{-11mm}

\begin{equation} \label{eqn:background-dnn-bw-weight-grad}
    	\frac{\partial E}{\partial W\textsuperscript{l}} \ (or \ \delta W\textsuperscript{l}) = X\textsuperscript{(l)} \ (\delta H\textsuperscript{(l+1)})\textsuperscript{T}, \
    	W\textsuperscript{l} = W\textsuperscript{l} - \eta * \frac{\partial E}{\partial W\textsuperscript{l}} \\
\end{equation}

\subsection{Using Crossbars for Training}

The most computationally intensive DNN layers that are typical targets for acceleration are the \textit{fully-connected} layers and the \textit{convolutional} layers.
We use fully-connected layers as an example to show how ReRAM crossbars can be used to accelerate DNN training workloads.

\subsubsection{Overview of Fully Connected (FC) Layers}

\begin{figure}[t]
  \centering
  \includegraphics[width=\columnwidth]{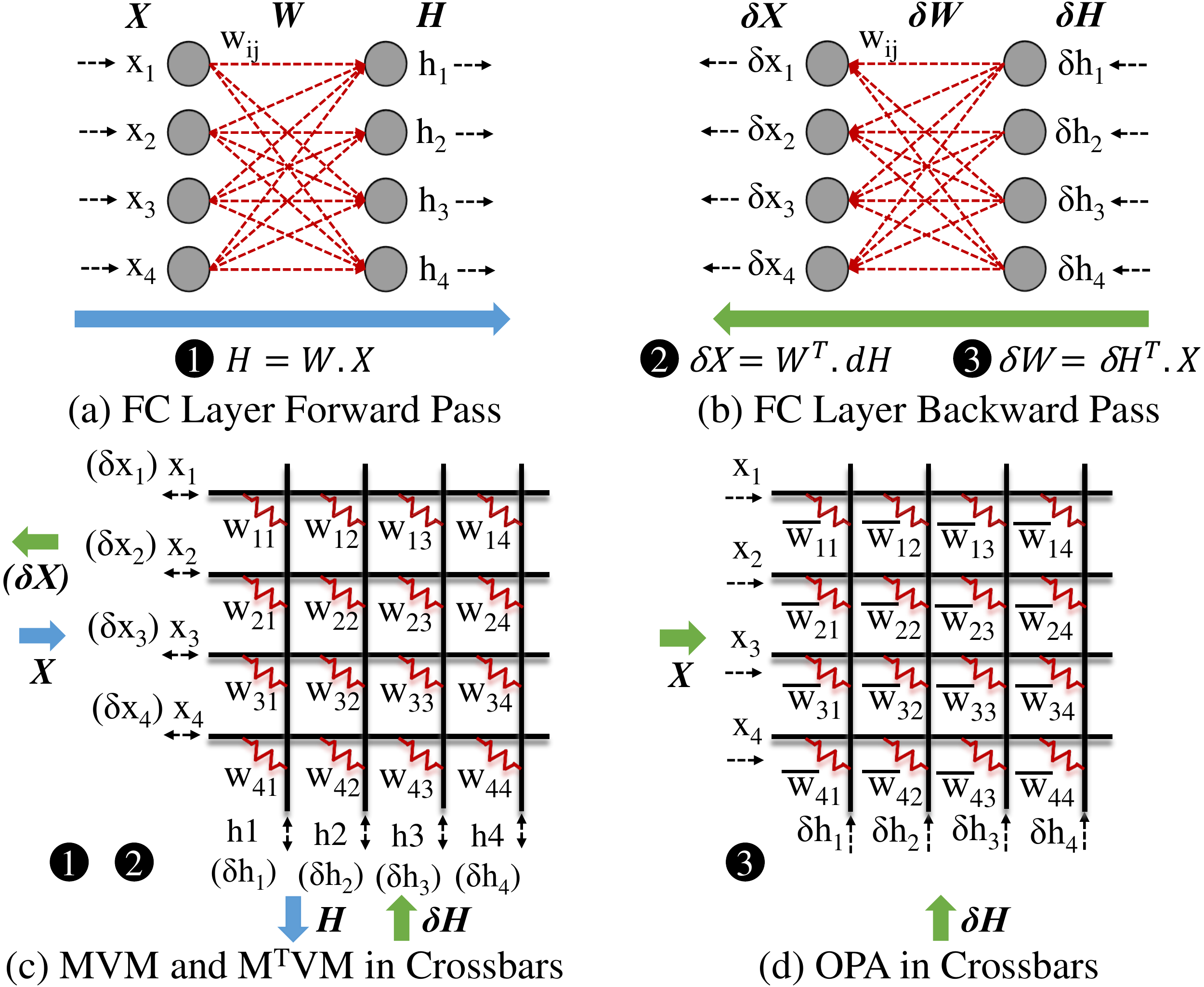}
  \shrinkBeforeCaption
  \caption{FC Layer Matrix Operations in Crossbars} \label{fig:mlp_crossbar}
  \shrinkAfterCaption
\end{figure}

Figures~\ref{fig:mlp_crossbar}(a) and~(b) illustrate the operations involved during training in a FC layer.
The training involves three types of matrix operations: \ding{182} \textit{activation}, \ding{183} \textit{layer gradients}, and \ding{184} \textit{weight gradients}.
Activation corresponds to an MVM operation with the weight matrix ($W$), as shown in Equation~\ref{eqn:background-dnn-fw}.
Layer gradients correspond to an MVM operation with the transpose of the weight matrix (hereon denoted as \MTVM), as shown in Equation~\ref{eqn:background-dnn-bw-layer-grad}.
Weight gradients correspond to an outer product operation, the result of which is accumulated to the weight matrix based on the learning rate ($\eta$), as shown in Equation~\ref{eqn:background-dnn-bw-weight-grad}.
Therefore weight gradients and updates together can be viewed as an Outer Product Accumulate (OPA) operation on the weight matrix.

\begin{figure*}[t]
  \centering
  \includegraphics[width=\textwidth]{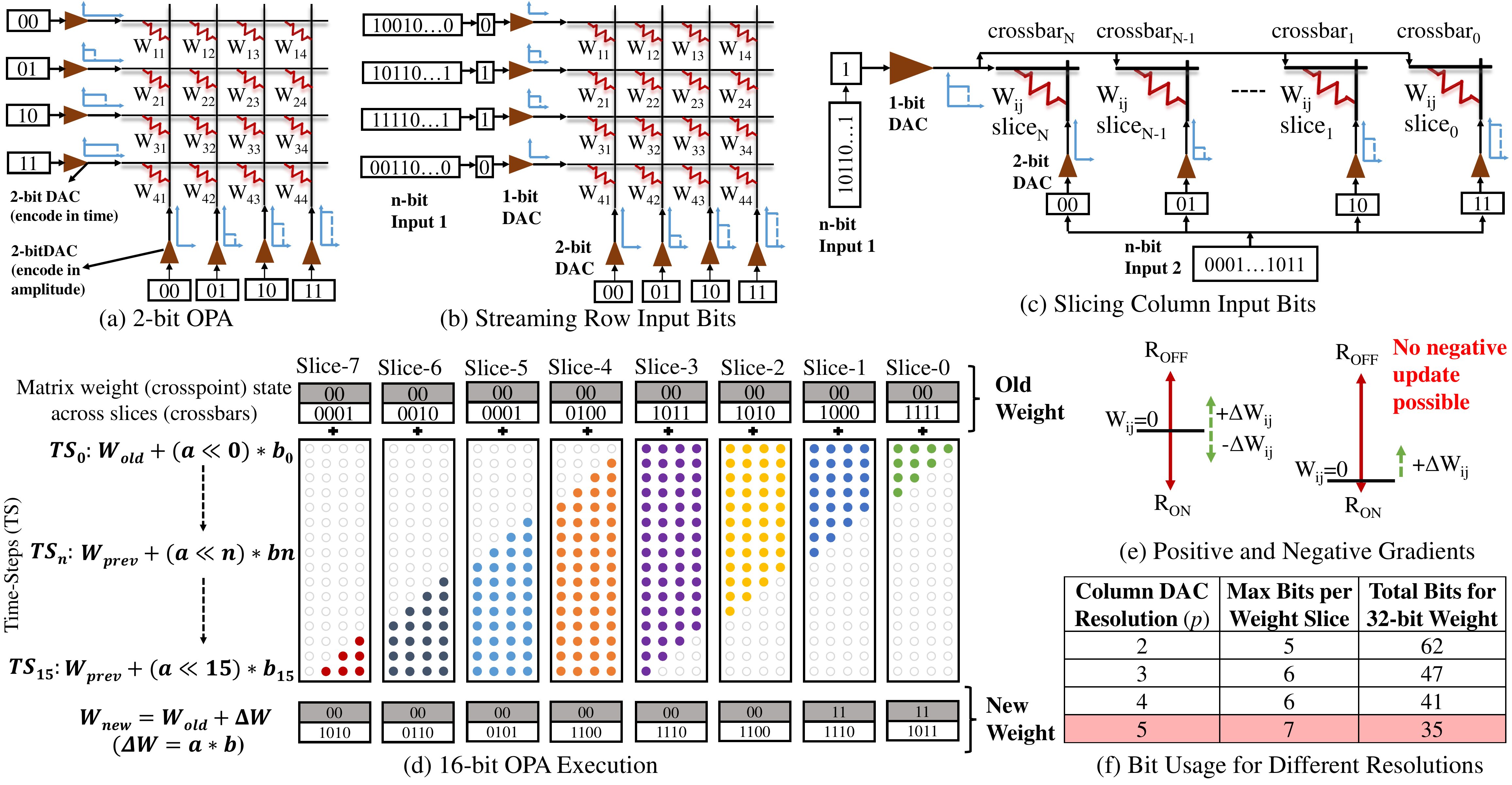}
  \shrinkBeforeCaption
  \caption{Bit Slicing OPA to Enhance its Precision}\label{fig:slice_op}
  \shrinkAfterCaption
\end{figure*}

\subsubsection{Activation and Layer Gradients in Crossbars}

Figure~\ref{fig:mlp_crossbar}(c) shows how a ReRAM crossbar can be used to compute activation and layer gradients.
The weights of the matrix ($W$) are stored in the crossbar cells as the conductance state~\cite{Hu2018MNIST}.
The MVM operation is realized by applying the input vector ($X$) as voltages on the rows of crossbar.
Subsequently, the output vector ($H$) is obtained as currents from the columns.
The \MTVM operation is realized by applying the input vector ($\delta H$) as voltages on the columns of the crossbar.
Subsequently, the output vector ($\delta X$) is obtained as currents from the rows.

Both MVM and \MTVM operations execute $O(n\textsuperscript{2})$ multiply-and-accumulate operations in one computational step in the analog domain ($n$ is the crossbar size).
Therefore, ReRAM crossbars can be leveraged to design highly efficient primitives for activation and layer gradient computations.
For this reason, they have been extensively considered for DNN inference~\cite{shafiee2016isaac,chi2016prime,liu2015reno} and training~\cite{cheng2017time,song2017pipelayer} accelerators.

\subsubsection{Weight Gradients and Updates in Crossbars}\label{sec:mlp-wg}

Figure~\ref{fig:mlp_crossbar}(d) shows how a ReRAM crossbar can be used to compute weight gradients.
The OPA operation can be realized by applying the inputs ($X$ and $\delta H$) as voltages on the crossbar's rows and columns, respectively.
The change ($\overline{w}\textsubscript{i}\textsubscript{j}-w\textsubscript{i}\textsubscript{j}$) in the value stored at a cross-point $(i,j)$ is equal to the product of the voltage on row $i$ and column $j$ (details in Section~\ref{sec:op-precision}).
Therefore, the outer product operation in the crossbar is naturally fused with the weight matrix accumulate operation.

The OPA operation executes $O(n\textsuperscript{2})$ multiply-and-accumulate operations in one computational step in the analog domain.
It avoids serial reads and writes to ReRAM crossbar cells, which is important because reads and writes have orders of magnitude higher cost (energy and latency) than in-crossbar computations (MVM, \MTVM, OPA).
Therefore, ReRAM crossbars can be leveraged to design highly efficient primitives for weight gradient computation and weight update.

The aforementioned technique has been demonstrated with low-precision inputs/outputs (2-4~bits) and weights (2-5~bits) on the SGD training algorithm for FC layers only~\cite{marinella2018multiscale,narayanan2017toward}.
In this paper, we enhance the technique with architecture support to increase its precision and cater to a multiple training algorithms and different layer types.

%% file: sec/03-op-precision.tex
\section{Enhancing ReRAM-based OPA Precision}\label{sec:op-precision}

DNN workloads require $16$ to $32$ bits of precision for training~\cite{micikevicius2017mixed, wu2018training}.
However, input digital-to-analog converters (DACs), crossbar cells, and output ADCs cannot support such levels of precision due to technology limitations and/or energy considerations.
For this reason, accelerators that use ReRAM crossbars for MVM/\MTVM operations typically achieve the required precision with bit-slicing~\cite{shafiee2016isaac}, where matrix bits are sliced across the cells of multiple crossbars, input bits are streamed at the crossbar rows/columns, and shift-and-add logic is used to combine the output bits at each column/row across crossbars (slices).

Bit-slicing matrices to also support OPA operations is different because both the rows and columns are simultaneously applied as inputs and the outputs are the crossbar cells themselves.
Moreover, bit-slicing for OPA operations presents additional constraints for the choice of bit distribution across slices.
This section describes our technique for bit-slicing the OPA operation (Section~\ref{sec:bit-slice-op}), and discusses the constraints it adds to the choice of bit distribution and how we address them (Sections~~\ref{sec:overflow-carry} to~\ref{sec:heterogeneous-slicing}).

\subsection{Bit Slicing the OPA Operation}\label{sec:bit-slice-op}

Figure~\ref{fig:slice_op}(a) illustrates how the OPA operation is performed when 2-bit inputs are applied at the rows and the columns.
The digital row input is encoded in the time-domain using pulse-width modulation.
The digital column input is encoded in the amplitude-domain using pulse-amplitude modulation.
Both pulse-width and pulse-amplitude modulations can be implemented using DACs.
The weight change in a cell depends on the duration and the amplitude of the pulses applied on the corresponding row and column respectively, thereby realizing an analog OPA operation~\cite{marinella2018multiscale,narayanan2017toward}.

To perform an OPA operation with 16-bit inputs, naively increasing the DAC resolution is infeasible because DAC power consumption grows rapidly with resolution ($N$) as:
\begin{equation} \label{eqn:arch-dac-power}
    	P\textsubscript{DAC} = \beta (2^N/N+1)V\textsuperscript{2}f\textsubscript{clk} \text{~\cite{kim2018input}}
\end{equation}
Instead, we propose an architectural scheme to realize a 16-bit OPA operation by bit-streaming the row input bits, bit-slicing the column input bits, and bit-slicing the matrix weights across multiple crossbars.

Figure~\ref{fig:slice_op}(b) illustrates how we stream row input bits, $m$ bits at a time over $16/m$ cycles.
Meanwhile column input bits are left-shifted by $m$-bits every cycle.
Since the number of cycles decrease linearly with $m$ while the cycle duration increases exponentially with $m$ due to pulse-width modulation of row input, we choose $m=1$ to minimize total latency.
Using $m=1$ also means that the row DACs are just inverters, thereby having low power consumption.

Figure~\ref{fig:slice_op}(c) shows how we slice column input bits across crossbars.
Only one weight $W\textsubscript{ij}$ is shown for clarity.
In each cycle, the left-shifted column input is divided into chunks of $p$ bits ($p=2$ in this example) and each chunk is applied to the corresponding crossbar.

Figure~\ref{fig:slice_op}(d) illustrates the steps for a 16-bit$\times$16-bit OPA operation at one crosspoint in the crossbar, resulting in a 32-bit output value for each matrix weight.
It puts together the bit-streaming of the row input vector \textbf{$b$} and bit-slicing of the column input vector \textbf{$a$} with $p=4$.
Each dot represents a partial product ($a\textsubscript{n}.b\textsubscript{n})$, and the color corresponds to a specific weight slice (crossbar).
Thus, the net accumulation to a slice is the result of all partial products of the specific color.
The updated weight after a time step $T\textsubscript{n}$ can be expressed as:

\vspace{-4mm}
\begin{equation} \label{eqn:arch-op-overall}
    	W\textsubscript{updated} = W\textsubscript{old} + \sum_{n=0}^{n} (a << n) * b\textsubscript{n}
\end{equation}

Crossbars store data in unsigned form.
To enable positive and negative weight updates ($\delta W$), we represent inputs in the signed magnitude representation.
To enable a symmetric representation of positive and negative weight updates, we bias each device such that, a zero weight ($W\textsubscript{ij}$) is represented by the memory state $(R\textsubscript{ON}+R\textsubscript{OFF})/2$, as shown in Figure~\ref{fig:slice_op}(e).
Hence, the signed magnitude computation and biased data representation enable both positive and negative updates to weights. 
This is important as both polarities of updates are equally important in DNN training.
Such a biased-representation can be implemented by adding an extra column per crossbar ($128$ rows, $128$ columns) with minimal area/energy cost~\cite{truong2014new}.

\subsection{Bits to Handle Overflow}\label{sec:overflow-carry}

For MVM/\MTVM, the matrix weights are inputs to the operation and they do not change.
In contrast, for OPA, the matrix weights are accumulated with the values resulting from the outer product.
As a result, the weight slice stored in a crossbar cell may overflow, either from multiple accumulations within one OPA or over multiple OPAs.
We handle this overflow by provisioning weight slices with additional bits to store the carry (shaded bits shown in Figure~\ref{fig:slice_op}(d)).

Propagating carry bits to other slices would require serial reads and writes which incur high overhead.
For this reason, we do not propagate the carry bits immediately.
Instead, they are kept in the slice and participate in future MVM/\MTVM and OPA operations on the crossbar.

The carry bits cannot be kept in the weight slice indefinitely because eventually the weight slice may get saturated i.e. crossbar cell at maximum/minimum state for positive/negative update.
Saturation is detrimental for \textit{trainability} (desirable loss reduction during training) because it freezes training progress due to the absence of weight change.
For this reason, we employ a periodic \textit{Carry Resolution Step} (CRS) which executes infrequently to perform carry propagation using serial reads and writes.
We evaluate the impact of the number of bits provisioned per slice and the CRS frequency on saturation and accuracy in Section~\ref{sec:eval-bits-crs}.

\subsection{Number of Slices vs. Bits Per Slice}

When slicing matrix bits across multiple crossbars, there is a tradeoff between the number of slices and the number of bits per cell in each slice.
MVM operations favor using more slices and fewer bits per slice.
The reason is that energy increases linearly with the number of crossbars, and non-linearly with the precision of a crossbar due to the increase in ADC precision required to support it.
Therefore, using more slices with fewer bits each is better for energy consumption.

In contrast, OPA favors having fewer slices with more bits per slice.
The reason is that OPA introduces carry bits to each slice and having more slices with fewer bits each increases the overhead from the carry bits.
For example, Figure~\ref{fig:slice_op}(f) shows that with 2 bits per slice, 62 total bits are required to represent the 32-bit weight while capturing the carry bits adequately.

To strike a balance, we choose $p=4$, since $p>4$ requires a device precision that exceeds ReRAM technology limits~\cite{Hu2018MNIST}.
A 4-bit DAC resolution is feasible because DAC power does not increase rapidly at low resolution (Equation~\ref{eqn:arch-dac-power}).
By choosing $p=4$, our MVM/\MTVM operations consume more energy than other ReRAM-based accelerators.
However, our more energy efficient OPA operations compensate because they avoid the need for expensive serial reads and writes.

\subsection{Heterogeneous Weight Slicing}\label{sec:heterogeneous-slicing}

MVM operations favor homogeneous bit-slicing.
Increasing the precision of one slice while decreasing the precision of another is always an unfavorable tradeoff because energy increases nonlinearly with the precision of a crossbar.
In contrast, for OPA operations where crossbar values change, provisioning more bits for slices that experience more weight updates helps reduce the frequency of saturation, thereby ensuring trainability while keeping the frequency of CRS low.

\begin{figure}[t]
  \centering
  \includegraphics[width=\columnwidth]{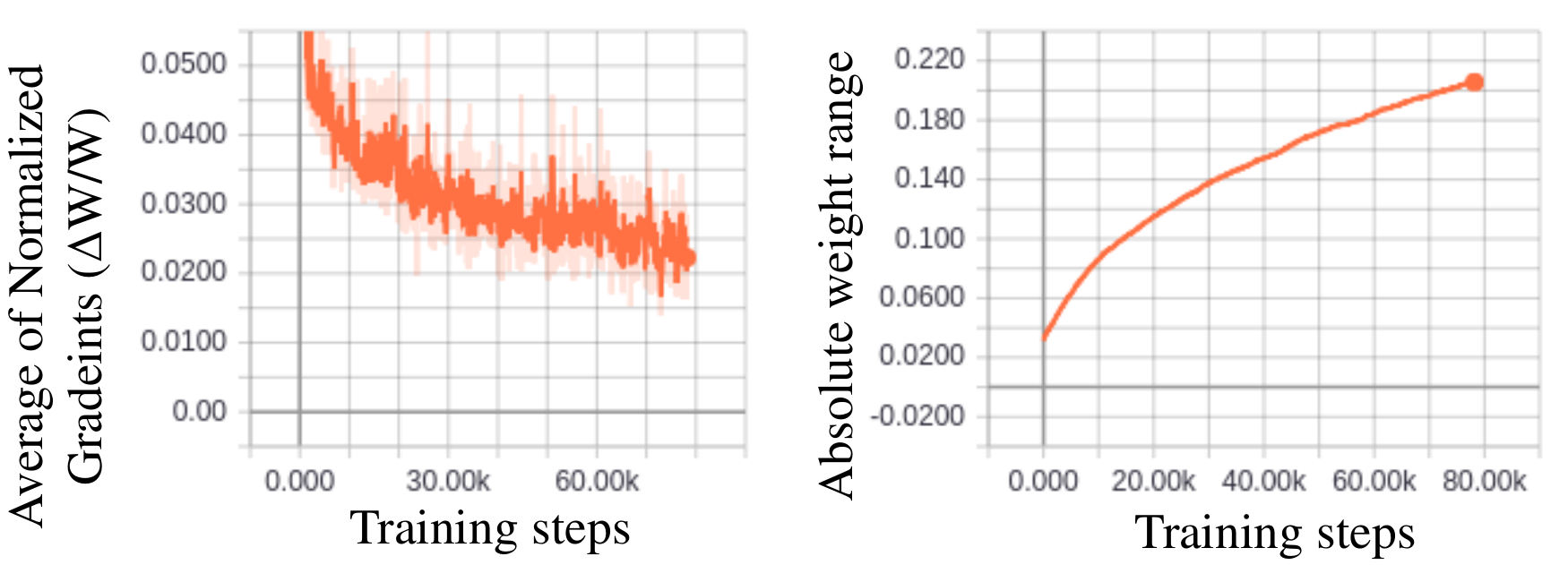}
  \shrinkBeforeCaption
  \caption{Weight Gradients across Training Steps} \label{fig:grad-range}
  \shrinkAfterCaption
\end{figure}

Heterogeneous weight slicing provisions more bits for matrix slices that change more frequently.
The frequency of change is impacted by two factors: OPA asymmetry and the small weight gradient range in DNNs.
OPA asymmetry is illustrated in Figure~\ref{fig:slice_op}(d) where the central slices receive more partial products (dots) than the edge slices, which motivates increasing precision for the central slices.
Small weight gradient range is shown in Figure~\ref{fig:grad-range} where weight updates form a very small fraction ($2\%-5\%$) of the overall weight range for $>=95\%$ of training steps, which motivates increasing precision of the lower slices.
We evaluate the impact of heterogeneous weight slicing on energy and accuracy in Section~\ref{sec:eval-heterogeneous-slicing}.

%% file: sec/04-mcu.tex
\section{Matrix Computation Unit (MCU)}\label{sec:mcu}

The techniques described in Section~\ref{sec:op-precision} are incorporated into a Matrix Computation Unit (MCU) for DNN training accelerators.
This section first describes the MCU's organization (Section~\ref{sec:mcu-organization}).
It then describes the three variants of the MCU optimized for
    SGD (Section~\ref{sec:arch-sgd}),
    mini-batch SGD (Section~\ref{sec:arch-mini-batch}),
    and mini-batch SGD with large batches (Section~\ref{sec:arch-mini-batch-large}).

\subsection{MCU Organization}\label{sec:mcu-organization}

\begin{figure}[t]
  \centering
  \includegraphics[width=0.95\columnwidth]{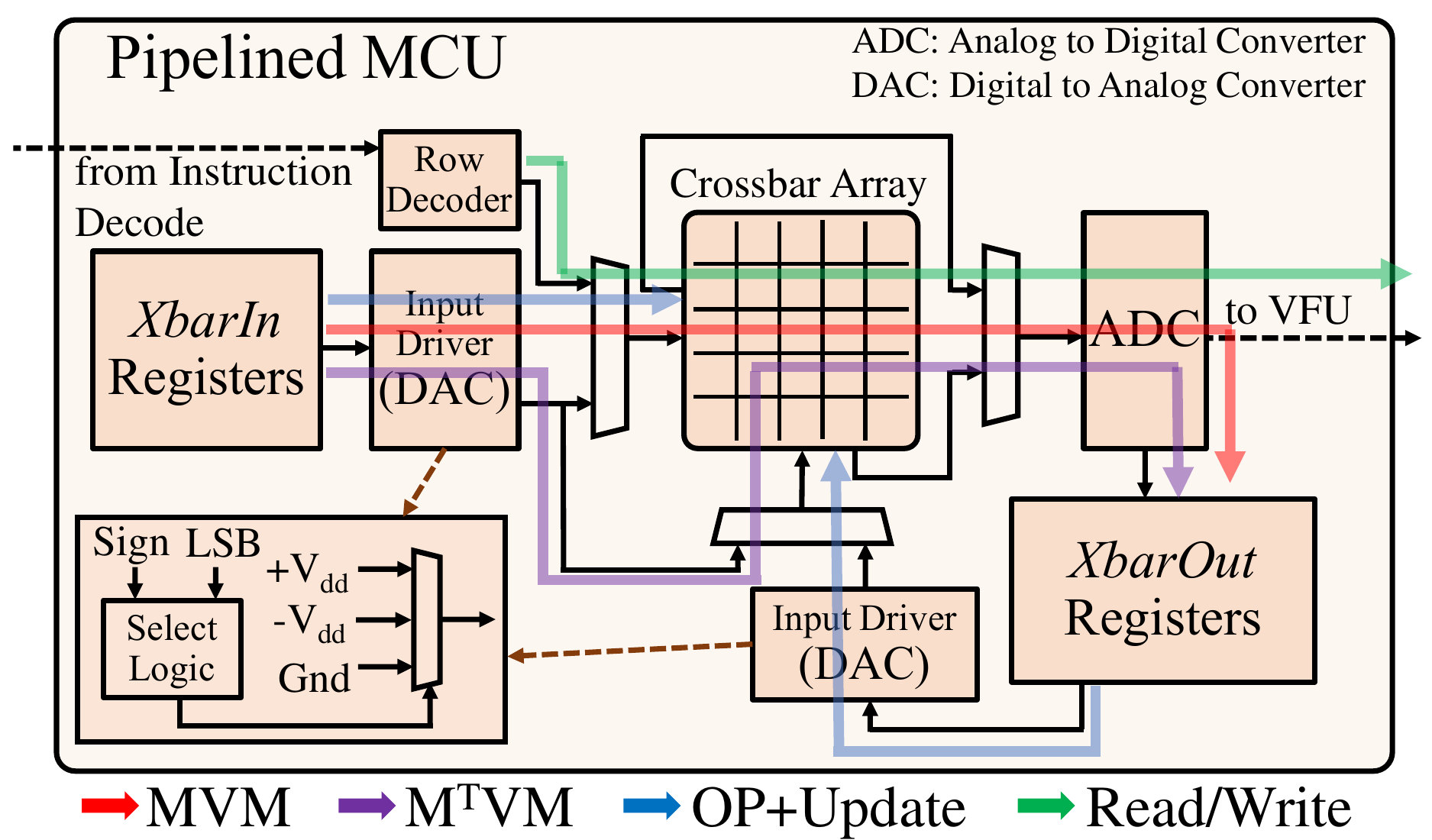}
  \vspace{-3mm}
  \caption{Matrix Computation Unit}\label{fig:arch-mcu}
  \vspace{-2mm}
\end{figure}

Figure~\ref{fig:arch-mcu} illustrates the organization of the MCU.
Performing an MVM operation with the MCU is illustrated by the red arrow.
Digital inputs stored in the \textit{XBarIn} registers are fed to the crossbar rows through the \textit{Input Driver}.
The output currents from the crossbar columns are then then converted to digital values using \textit{ADC} and stored in the \textit{XBarOut} registers.

Performing a \MTVM operation in the MCU is illustrated by the purple arrow in Figure~\ref{fig:arch-mcu}.
The key difference compared to the MVM operation is the addition of multiplexers to supply inputs to crossbar columns instead of rows and to read outputs from crossbar rows instead of columns.

MVM and \MTVM operations require $16$ to $32$ bits of precision for training~\cite{chen2014dadiannao}.
We use 16-bit fixed-point representation for input/output data and 32-bit fixed-point representation for weight data which ensures sufficient precision~\cite{micikevicius2017mixed}.

Performing an OPA operation in the MCU is illustrated by the blue arrow in Figure~\ref{fig:arch-mcu}.
Digital inputs stored in the \textit{XBarIn} registers are fed to the crossbar rows through the \textit{Input Driver}.
Digital inputs stored in the \textit{XBarOut} registers are fed to the crossbar columns through the \textit{Input Driver}.
The effect of this operation is that the outer product of the input vectors is accumulated to the matrix stored in the crossbar.
To support positive and negative inputs, the input drivers in Figure~\ref{fig:arch-mcu} use the sign bit (MSB) to drive the crossbar rows and columns with positive or negative voltages.

\subsection{Variant \#1 for SGD Acceleration} \label{sec:arch-sgd}

SGD-based training performs example-wise gradient descent.
First, an input example performs a forward pass (MVM) to generate activations - $H\textsuperscript{l}$. 
Next, the error computed with respect to the activation of the output layer is back propagated (\MTVM) to compute the layer gradients - $\delta X\textsuperscript{l}$.
Finally, the activations and layer gradients are used to update (OPA) the weight matrix - $W\textsuperscript{l}$, before the next input example is supplied.

\begin{table}[t]
\centering
    \caption{Dataflow for SGD}\label{tab:sgd-dataflow}
    \vspace{-4mm}
    \centering
    \resizebox{\columnwidth}{!}{
        \input{fig/04-mcu/sgd-dataflow.tex}
    }
    \vspace{-4mm}
\end{table}

Table~\ref{tab:sgd-dataflow} illustrates the logical execution of matrix operations in three MCUs for a three-layer DNN with an input example $a0$.
Each time step shows the operations executed on each MCU and their inputs/outputs.
For example, at time step $0$, MCU0 performs an MVM operation on input $a0$ to compute the output $a1$. 
The illustration assumes that each layer maps on one MCU and does not show the interleaved nonlinear operations for clarity.
For a layer size larger than one MCU capacity ($128\times128$ matrix), the layer is partitioned across multiple MCUs (see Section~\ref{sec:compiler}).

Variant \#1 of the MCU uses a single crossbar to perform all three matrix operations: MVM, \MTVM, and OPA.
This variant is suitable for SGD because, as shown in Table~\ref{tab:sgd-dataflow}, the three matrix operations are data dependent and will never execute concurrently.
However, this variant creates structural hazards for mini-batch SGD as described in Section~\ref{sec:arch-mini-batch}.

\subsection{Variant \#2 for Mini-Batch SGD Acceleration} \label{sec:arch-mini-batch}

Mini-batch SGD performs batch-wise gradient descent.
Like SGD, each input performs MVM, \MTVM, and OPA to compute activations, layer gradients, and weight gradients/updates, respectively.
However, the weight update is only reflected at the end of a batch to be used by the inputs of the next batch.

\begin{table}[t]
\centering
    \caption{Dataflow for Mini-Batch SGD}\label{tab:mini-batch-dataflow}
    \vspace{-4mm}
    \centering
    \resizebox{0.95\columnwidth}{!}{
        \input{fig/04-mcu/mini-batch-dataflow.tex}
    }
    \vspace{-4mm}
\end{table}

Table~\ref{tab:mini-batch-dataflow} illustrates the logical execution of matrix operations for a batch of five inputs, where $a\textsubscript{n}m$ refers to the m\textsuperscript{th} activation of n\textsuperscript{th} input.
MVM operations can be executed for multiple input examples concurrently in a pipelined fashion (\textbf{MVM} (a\textsubscript{1}0) (a\textsubscript{1}1), \textbf{MVM} (a\textsubscript{0}1) (a\textsubscript{0}2) in Table~\ref{tab:mini-batch-dataflow}).
Additionally, the MVM and \MTVM operations for different inputs in the batch can also execute in parallel during the same timestep, provided that there is no structural hazard on the MCU.
The desire to eliminate such structural hazards motivates Variant \#2.

Variant \#2 of the MCU eliminates structural hazards in mini-batch SGD by storing two copies of the matrix on different crossbars, enabling the MCU to perform MVM and \MTVM in parallel.
This replication improves the \textit{energy-delay product} for a batch.
With $<2\times$ increase in area, we improve the batch latency by $O(L)$, where $L$ is the number of layers.
The ISA instruction for performing MVM/\MTVM (Section~\ref{sec:isa}) is designed to enable the compiler (Section~\ref{sec:compiler}) to schedule these two operations in parallel on the same MCU.

The OPA operations are executed at the end of the mini-batch (steps 9-12 in Table~\ref{tab:mini-batch-dataflow}) to reflect the weight updates for the entire batch.
These OPA operations require that the vectors involved are saved until then.
Variant \#2 saves these vectors in shared memory.
However, if the batches are large, this approach puts too much stress on the shared memory which motivates Vaiant \#3 (Section~\ref{sec:arch-mini-batch-large}).

\subsection{Variant \#3 for Mini-Batch SGD with Large Batches} \label{sec:arch-mini-batch-large}

For mini-batch SGD with very large batch sizes, saving the vectors in shared memory requires large shared memory size which degrades storage density.
Variant \#3 alleviates the pressure shared memory size by maintaining three copies of each crossbar.
The first two copies enable performing MVM and \MTVM in parallel, similar to Variant \#2.
The third copy is used to perform the OPA operation eagerly, as soon as its vector operands are available, without changing the matrices being used by the MVM and \MTVM operations.

Performing OPA eagerly avoids saving vectors until the end, reducing the pressure on the shared memory.
However, using a third crossbar for OPA requires serial reads and writes to commit the weight updates to the first and the second crossbars for MVM and \MTVM in the next batch.
Section~\ref{sec:panther-2x-3x-comparision} discusses the impact of these design choices.

%% file: fig/04-mcu/sgd-dataflow.tex
\begin{tabular}{|c|l|l|l|}
\hline
\textbf{\begin{tabular}[c]{@{}c@{}}Time \\ Step\end{tabular}} & \textbf{MCU0 (Layer1)}          & \textbf{MCU1 (Layer2)} & \textbf{MCU2 (Layer3)} \\ \hline
0                                                             & \textbf{MVM} (a0) (a1)          &                       &                       \\ \hline
1                                                             &                                 & \textbf{MVM} (a1) (a2)         &                       \\ \hline
2                                                             &                                 &                       & \textbf{MVM} (a2) (a3)         \\ \hline
3                                                             &                                 &                       & \textbf{\MTVM} ($\delta$h3) ($\delta$h2)    \\ \hline
4                                                             &                                 & \textbf{\MTVM} ($\delta$h2), ($\delta$h1)   & \textbf{OP} (a2 , $\delta$h3) ($\nabla$W3)   \\ \hline
5                                                             & \textbf{OP} (a0, $\delta$h1) ($\nabla$W1)                                & \textbf{OP} (a1, $\delta$h2) ($\nabla$W2)    &                       \\ \hline
\end{tabular}

%% file: fig/04-mcu/mini-batch-dataflow.tex

\begin{tabular}{|c|l|l|l|}
\hline
\textbf{\begin{tabular}[c]{@{}c@{}}Time \\ Step\end{tabular}} & \textbf{MCU0 (Layer1)}          & \textbf{MCU1 (Layer2)}          & \textbf{MCU2 (Layer3)}          \\ \hline
\multirow{2}{*}{\textbf{0}}                                   & \textbf{MVM} (a\textsubscript{0}0) (a\textsubscript{0}1)       & \textbf{}                      & \textbf{}                      \\ \cline{2-4} 
                                                              & \textbf{}                      & \textbf{}                      & \textbf{}                      \\ \hline
\multirow{2}{*}{\textbf{1}}                                   & \textbf{MVM} (a\textsubscript{1}0) (a\textsubscript{1}1)       & \textbf{MVM} (a\textsubscript{0}1) (a\textsubscript{0}2)       & \textbf{}                      \\ \cline{2-4} 
                                                              & \textbf{}                      & \textbf{}                      & \textbf{}                      \\ \hline
\multirow{2}{*}{\textbf{2}}                                   & \textbf{MVM} (a\textsubscript{2}0) (a\textsubscript{2}1)       & \textbf{MVM} (a\textsubscript{1}1) (a\textsubscript{1}2)       & \textbf{MVM} (a\textsubscript{0}2) (a\textsubscript{0}3)       \\ \cline{2-4} 
                                                              & \textbf{}                      & \textbf{}                      & \textbf{}                      \\ \hline
\multirow{2}{*}{\textbf{3}}                                   & \textbf{MVM} (a\textsubscript{3}0) (a\textsubscript{3}1)       & \textbf{MVM} (a\textsubscript{2}1) (a\textsubscript{2}2)       & \textbf{MVM} (a\textsubscript{1}2) (a\textsubscript{1}3)       \\ \cline{2-4} 
                                                              & \textbf{}                      &                               & \textbf{\MTVM} ($\delta$h\textsubscript{0}3) ($\delta$h\textsubscript{0}2)                      \\ \hline
\multirow{2}{*}{\textbf{4}}                                   & \textbf{MVM} (a\textsubscript{4}0) (a\textsubscript{4}1)       & \textbf{MVM} (a\textsubscript{3}1) (a\textsubscript{3}2)       & \textbf{MVM} (a\textsubscript{2}2) (a\textsubscript{2}3)       \\ \cline{2-4} 
                                                              &  & \textbf{\MTVM} ($\delta$h\textsubscript{0}2), ($\delta$h\textsubscript{0}1)  & \textbf{\MTVM} ($\delta$h\textsubscript{1}3) ($\delta$h\textsubscript{1}2)                      \\ \hline
\multirow{2}{*}{\textbf{5}}                                   & \textbf{}                      & \textbf{MVM} (a\textsubscript{4}1) (a\textsubscript{4}2)       & \textbf{MVM} (a\textsubscript{3}2) (a\textsubscript{3}3)       \\ \cline{2-4} 
                                                              &  & \textbf{\MTVM} ($\delta$h\textsubscript{1}2), ($\delta$h\textsubscript{1}1)  & \textbf{\MTVM} ($\delta$h\textsubscript{2}3) ($\delta$h\textsubscript{2}2)                      \\ \hline
\multirow{2}{*}{\textbf{6}}                                   & \textbf{}                      & \textbf{}                      & \textbf{MVM} (a\textsubscript{4}2) (a\textsubscript{4}3)       \\ \cline{2-4} 
                                                              &  &  \textbf{\MTVM} ($\delta$h\textsubscript{2}2), ($\delta$h\textsubscript{2}1)  & \textbf{\MTVM} ($\delta$h\textsubscript{3}3) ($\delta$h\textsubscript{3}2)                      \\ \hline
\multirow{2}{*}{\textbf{7}}                                   & \textbf{}                      & \textbf{}                      & \textbf{}                      \\ \cline{2-4} 
                                                              &  &  \textbf{\MTVM} ($\delta$h\textsubscript{3}2), ($\delta$h\textsubscript{3}1)  & \textbf{\MTVM} ($\delta$h\textsubscript{4}3) ($\delta$h\textsubscript{4}2)                      \\ \hline
\multirow{2}{*}{\textbf{8}}                                   & \textbf{}                      & \textbf{}                      & \textbf{}                      \\ \cline{2-4} 
                                                              &  & \textbf{\MTVM} ($\delta$h\textsubscript{4}2), ($\delta$h\textsubscript{4}1)                      & \textbf{}                      \\ \hline
\multirow{2}{*}{\textbf{9-12}}                                & \textbf{OP} (a\textsubscript{n}0, $\delta$h\textsubscript{n}1) ($\nabla$W\textsubscript{n}1) & \textbf{OP} (a\textsubscript{n}1, $\delta$h\textsubscript{n}2) ($\nabla$W\textsubscript{n}2) & \textbf{OP} (a\textsubscript{n}2, $\delta$h\textsubscript{n}3) ($\nabla$W\textsubscript{n}3) \\ \cline{2-4} 
                                                              & \multicolumn{3}{c|}{\textbf{Iterate for n=1 to 4}}                                               \\ \hline
\end{tabular}

%% file: sec/05-accelerator.tex
\section{Programmable Accelerator}\label{sec:accelerator}

The MCU described in Section~\ref{sec:mcu} can be integrated with prior ReRAM-based training accelerators~\cite{cheng2017time,song2017pipelayer} to improve their efficiency.
We develop a programmable training accelerator \additionx{named \reramop} to evaluate our design
\additionx{by extending the PUMA ReRAM-based inference accelerator~\cite{ankit2019puma}}.
This section describes \additionx{\reramop{}'s} organization (Section~\ref{sec:arch-micro}), ISA considerations (Section~\ref{sec:isa}), compiler support (Section~\ref{sec:compiler}), and an example of how to implement convolutional layers (Section~\ref{sec:cnn}).

\subsection{Accelerator Organization} \label{sec:arch-micro}

\begin{figure}[t]
  \centering
  \includegraphics[width=\columnwidth]{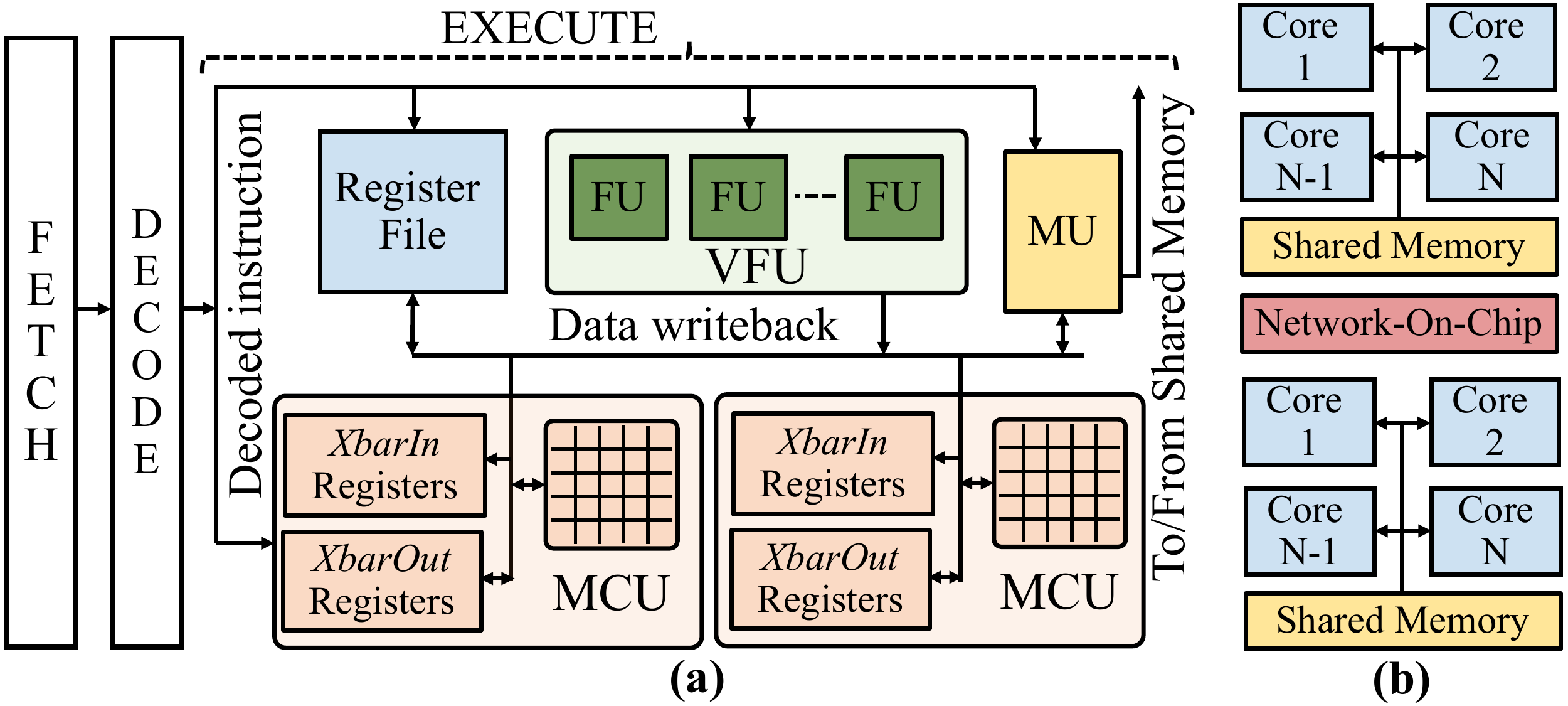}
  \shrinkBeforeCaption
  \caption{Architecture Overview} \label{fig:arch-core}
  \shrinkAfterCaption
\end{figure}

\additionx{\reramop} is a spatial architecture organized in three tiers: nodes, tiles, and cores.
A node consists of multiple tiles connected via an on-chip network, and a tile consists of multiple cores connected to a shared memory, as illustrated in Figure~\ref{fig:arch-core}(b).
A core consists of multiple MCUs for executing matrix operations, a digital CMOS-based vector functional unit (VFU) for executing arithmetic operations and non-linear functions, a register file, and a load/store memory unit.
A core also features an instruction execution pipeline making the accelerator ISA-programmable.
To support DNNs whose model storage exceeds a node's total MCU capacity, multiple nodes can be connected via an interconnect.
This organization is \additionx{similar to PUMA's~\cite{ankit2019puma}} and is not a contribution of this paper.
\additionx{The key distinction from PUMA is the MCU which supports \MTVM and OPA operations, not just MVM operations, as described in Section~\ref{sec:mcu}.}

\subsection{ISA Considerations}\label{sec:isa}

The \additionx{PUMA~\cite{ankit2019puma}} ISA includes \additionx{\textit{mvm} instructions executed by crossbars,} arithmetic/logic/nonlinear instructions executed by the VFU, load/store instructions to access shared memory, send/receive instructions to communicate with other tiles, and control flow instructions.
\additionx{We extend the PUMA ISA to} also include a \textit{mcu} instruction for executing all three matrix operations (MVM, \MTVM, OPA) on the MCU.

The \textit{mcu} instruction takes six 3-bit masks, where each mask corresponds to one of the MCUs on the core (up to six).
The three bits in the mask correspond to the three supported matrix operations (MVM, \MTVM, OPA).
If multiple bits are set, then the instruction executes the operations concurrently.
For example, if mask~0 is set to '110' and mask~1 is set to '011', then MCU~0 will execute MVM and \MTVM simultaneously and MCU~1 will execute \MTVM and OPA simultaneously.
\additionx{Hence, the incorporation of all three operations into a single instruction is important for being able to execute them concurrently.}
The \textit{mcu} instruction does not take source and destination operands since these are implied to by \textit{XBarIn} and \textit{XBarOut}.

The semantic of the OPA operation is that it takes effect at the end of the execution when a special \textit{halt} instruction is invoked.
This semantic allows the same code to work for any of the three MCU variants, making the choice of variant a microarchitectural consideration and the ISA agnostic to it.
The implementation of the OPA semantic on each of the variants is as follows.
Consider the case when all three bits of an MCU's mask are set.
In Variant \#1, MVM and \MTVM will be serialized on the same crossbar, while the operands of OPA will be saved to shared memory then applied to that crossbar when \textit{halt} is invoked.
In Variant \#2, MVM and \MTVM will be executed in parallel on the two crossbar copies, while the operands of OPA will be treated like in Variant \#1.
In Variant \#3, MVM and \MTVM will be executed in parallel on the first two crossbar copies, while the operands of OPA will be applied to the third crossbar.
The values of the third crossbar will then be copied to the first two crossbars when \textit{halt} is invoked.

\subsection{Compiler Support}\label{sec:compiler}

\additionx{The PUMA~\cite{ankit2019puma}} compiler provides a high-level programming interface in C++ that allows programmers to express models in terms of generic matrix and vector operations.
The compiler is implemented as a runtime library that builds a computational graph when the code is executed then compiles the graph to PUMA ISA code.
The compiler partitions matrices into sub-matrices and maps these sub-matrices to different MCUs, cores, and tiles.
It then maps the operations in the graph to different MCUs, cores, and tiles accordingly, inserting communication operations where necessary.
The compiler then linearizes the graph, creating an instruction sequence for each core.
It performs register allocation for each sequence, spilling registers to shared memory if necessary.
Finally, it generates ISA code for each core, collectively comprising a kernel that runs on the accelerator.

\additionx{
We make the following extensions to the PUMA compiler to support \reramop.
We extend the application programming interface (API) to allow programmers to define training matrices that support MVM, \MTVM{}, and OPA operations.
We extend the intermediate representation to represent these matrices and include them in the partitioning.
We also add an analysis and transformation pass for identifying MCU operations in the graph that can be fused and fusing them.
This pass fuses
}
MCU operations that do not have data dependences between them and that use different MCUs on the same core or use the same MCU but are different types of operations (MVM, \MTVM, OPA).
The fusing process is iterative because every time operations are fused, new dependences are introduced to the graph.
\additionx{
Finally, we extend the code generator to support the new \textit{mcu} ISA instruction.
}

Note that since the model weights are not updated until the \textit{halt} instruction at the end, the scope of a kernel is a single batch.
Multiple batches are executed by invoking the kernel multiple times on different input data.

\subsection{Implementing Convolutional Layers}\label{sec:cnn}

ReRAM-based OPA has one-to-one correspondence to the weight gradient/update operation for FC layers (discussed in Section~\ref{sec:mlp-wg}).
By integrating this technique into a programmable accelerator with compiler support, we enable the mapping of more complex layers on top of it such as convolutional layers.
This section describes how convolutional layers can be implemented in our accelerator.

\begin{figure}[t]
  \centering
  \includegraphics[width=1\columnwidth]{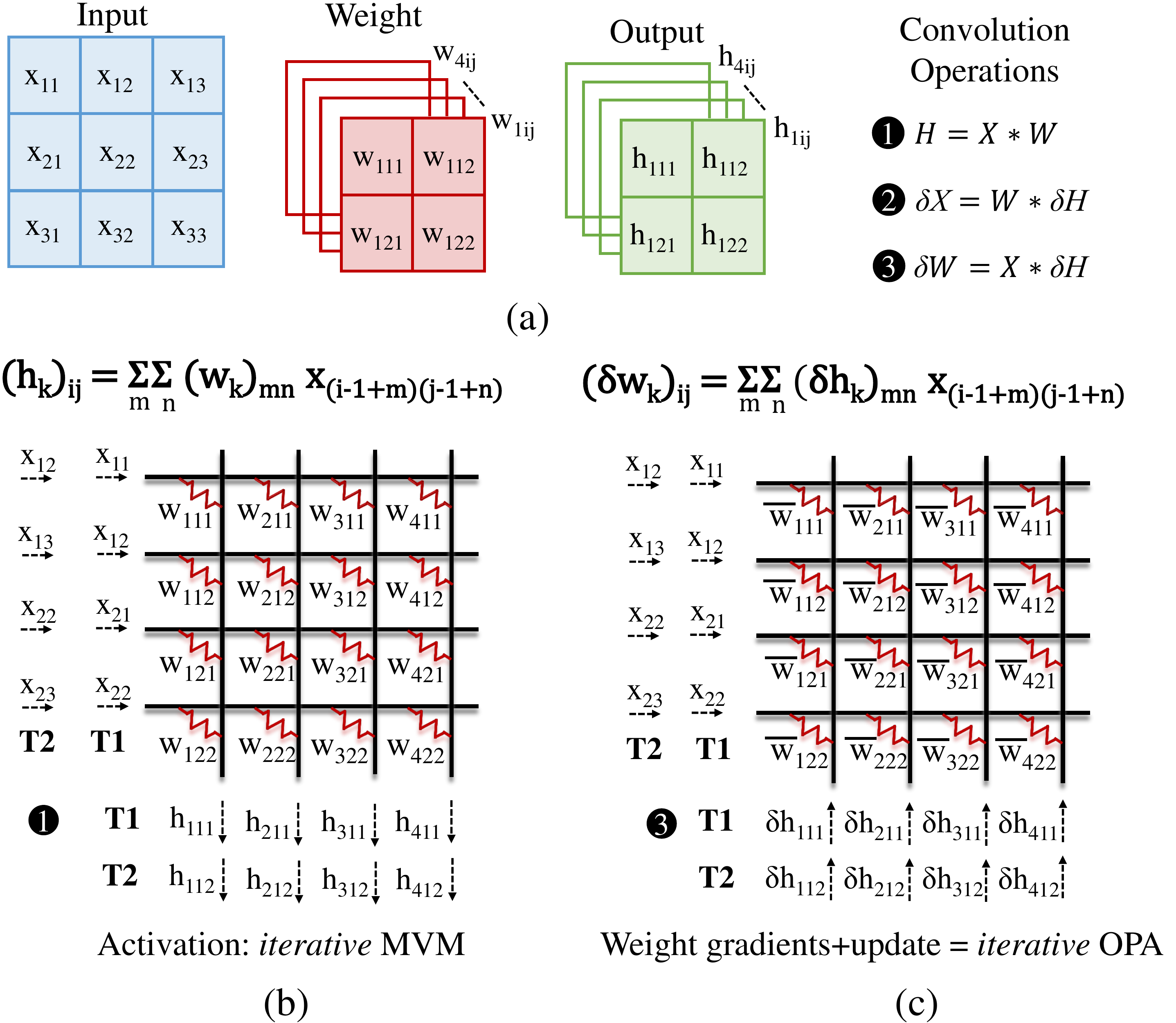}
  \vspace{-8mm}
  \caption{Convolutional Layer Matrix Operations in Crossbars} \label{fig:cnn_crossbar}
  \vspace{-3mm}
\end{figure}

Figure~\ref{fig:cnn_crossbar}(a) shows a typical convolution layer and the associated operations during training.
Like with FC layers, convolutional layers performs three types of matrix operations: \ding{182} \textit{activation}, \ding{183} \textit{layer gradients}, and \ding{184} \textit{weight gradients}.
Unlike FC layers, these operations are all convolutions ($*$).

\subsubsection{Activation and Layer Gradients}

Figure~\ref{fig:cnn_crossbar}(b) shows how the convolution operation for activation is implemented in the crossbar on top of the MVM primitive.
This approach is similar to that used in existing accelerators~\cite{song2017pipelayer}.
The crossbar stores the \textit{convolution kernel} in the form of linearized filters ($w\textsubscript{k}$), where each column corresponds to the weights associated with a specific output channel ($h\textsubscript{k}$).
The convolution operation to compute activations is implemented as an \textit{iterative} MVM operation.
An iteration is represented as a time step (T1/T2) in Figure~\ref{fig:cnn_crossbar}(b), and corresponds to a specific (i,j) pair.
A block of input features ($X$) is applied to the crossbar's rows as \textit{convolution data} in each iteration. 
In a similar manner, the convolution operation for layer gradients (not shown in the figure) is realized using iterative \MTVM.
The next layer's errors ($\delta H$) are used as the convolution data and flipped filters (vertically and horizontally) are used as the convolution kernel.

\subsubsection{Weight Gradients} \label{sec:cnn-wt-grad}

Figure~\ref{fig:cnn_crossbar}(c), shows our proposed technique for implementing the weight gradients convolution operation and weight update in the crossbar on top of the OPA primitive.
The weight gradient computation uses input features ($X$) as the convolution data and output feature's errors ($\delta H$) as the convolution kernel.
Each iteration is represented as a time step (T1, T2) in Figure~\ref{fig:cnn_crossbar}(c), and corresponds to a specific (i,j) pair.
On every iteration, the output feature's errors are applied on the columns, in a depth major order.
Simultaneously, by applying the portion of input features that generate the corresponding activations ($H$) on the rows, a \textit{partial convolution} is obtained between $X$ and $\delta H$.
Striding across the output feature's errors and input features for $n\textsuperscript{2}$ time steps, where $n$ is size of one output feature map, realizes the full convolution operation.
Convolutions for different output feature maps are performed in parallel across the crossbar's columns, using the same weight data layout as used in MVM and \MTVM operations.
To the best of our knowledge, our work is the first to formulate the weight gradients convolution operation in terms of outer products.

\subsubsection{Comparison with Other Accelerators}\label{sec:compare-pipelayer}

Existing ReRAM-based training accelerators such as PipeLayer \cite{song2017pipelayer} do not compute the weight gradient convolutions using outer products, but rather, they compute them using MVM operations.
This requires writing the convolution kernel ($\delta H$) on the crossbar because the convolution operation here uses \textit{non-stationary} data ($\delta H$) as the convolution kernel.
The drawback of this approach is that the latency and energy consumption of the serial reads and writes is very high, taking away from the overall efficiency provided by ReRAM-based MVMs.

%% file: sec/06-methodology.tex
\section{Methodology}\label{sec:methodology}

\begin{figure}[t]
  \centering
  \includegraphics[width=\columnwidth]{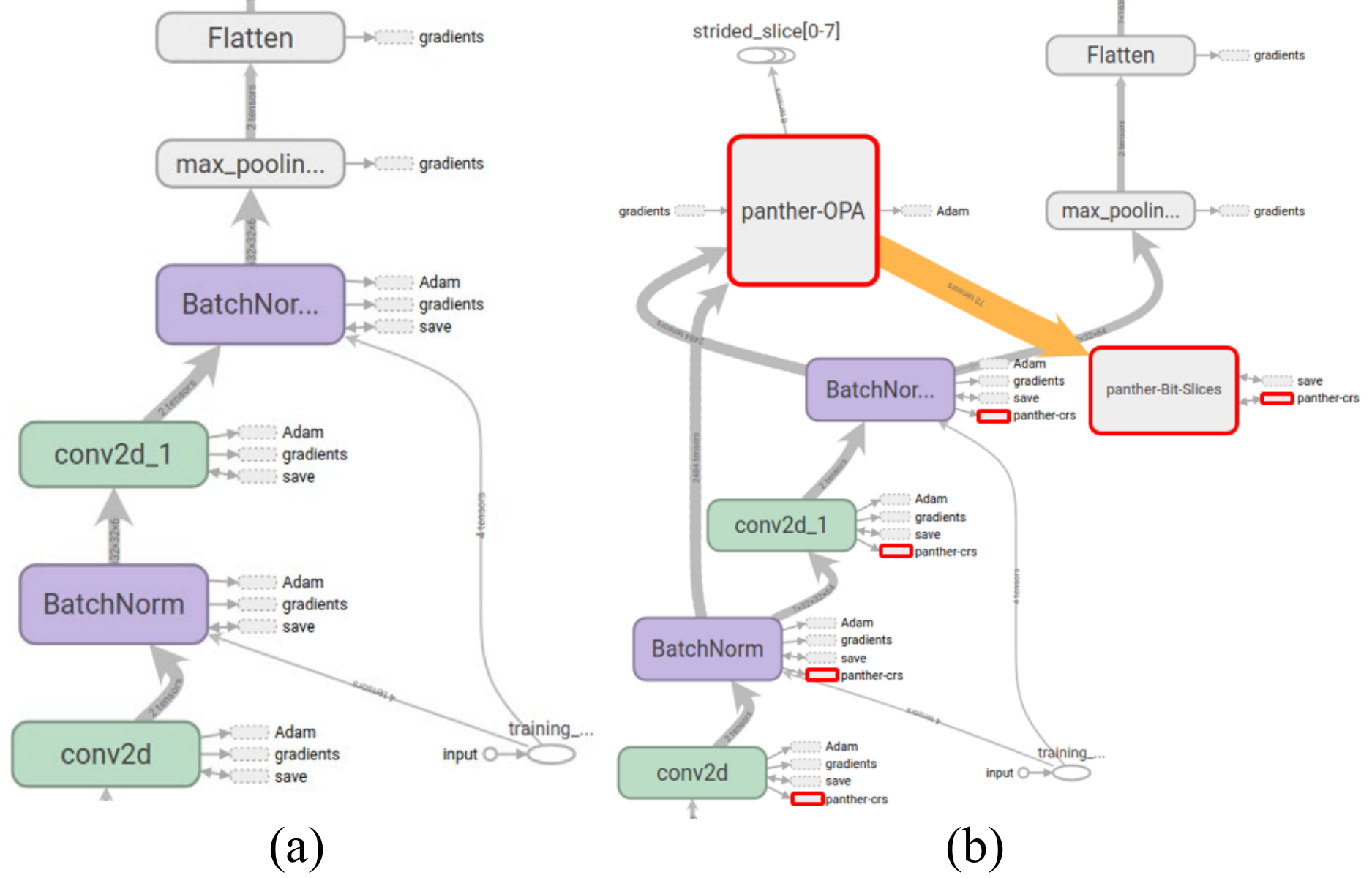}
  \shrinkBeforeCaption
  \caption{\addition{Computational graph obtained using TensorBoard for (a) example model (b) example model with PANTHER OPA}} \label{fig:tensorboard-graphs}
  \vspace{-4mm}
\end{figure}

\subsection {Architecture Simulator}

\addition{
We extend the PUMA~\cite{ankit2019puma} simulator to model the MCU unit and its associated instructions.
The PUMA simulator is a detailed cycle-level architecture simulator that runs applications compiled by the compiler, in order to evaluate the execution of benchmarks.
The simulator models all the necessary events that occur in an execution cycle, including compute, memory and NoC transactions.
To estimate power and timing of the CMOS digital logic components, their RTL implementations are synthesized to the IBM 32nm SOI technology library, and evaluated using the Synopsys Design Compiler.
For the on-chip SRAM memories, the power and timing estimates are obtained from Cacti 6.0.
Subsequently, the power and timing of each component are incorporated in the cycle-level simulator in order to estimate the energy consumption.

\textbf{MCU Modelling.}
Since the MCU is built with analog components and cannot be synthesized with publicly available libraries, we adopted the models from past works~\cite{shafiee2016isaac,marinella2018multiscale} and ADC survey~\cite{murmann2011adc}.
We use the ReRam crossbar array and sample-and-hold circuit models in ISAAC~\cite{shafiee2016isaac}.
We used capacitive DACs and Successive Approximation Register (SAR) ADCs.
The DAC area and power are estimated using the equations described in Saberi et al.~\cite{saberi2011analysis}.
The ADCs for different precisions namely 8-12 bits operating at a sampling frequency of $1GHz$ are obtained from the ADC survey~\cite{murmann2011adc}.
The ADC optimization technique in Newton~\cite{nag2018newton} is incorporated to avoid unnecessary ADC conversions.
}

\begin{table}[t]
\centering
    \caption{\addition{Summary of platforms}}\label{tab:hw-platforms}
    \vspace{-3mm}
    \centering
    \resizebox{0.9\columnwidth}{!}{
        \input{fig/06-methodology/hw-platform.tex}
    }
    \vspace{-4mm}
\end{table}

\begin{table}[t]
\centering
    \caption{\addition{Details of workloads}}\label{tab:workload}
    \vspace{-3mm}
    \centering
    \resizebox{1.0\columnwidth}{!}{
        \input{fig/06-methodology/workload-details.tex}
    }
    \vspace{-5mm}
\end{table}

\begin{figure*}
    \centering
    \includegraphics[width=\textwidth]{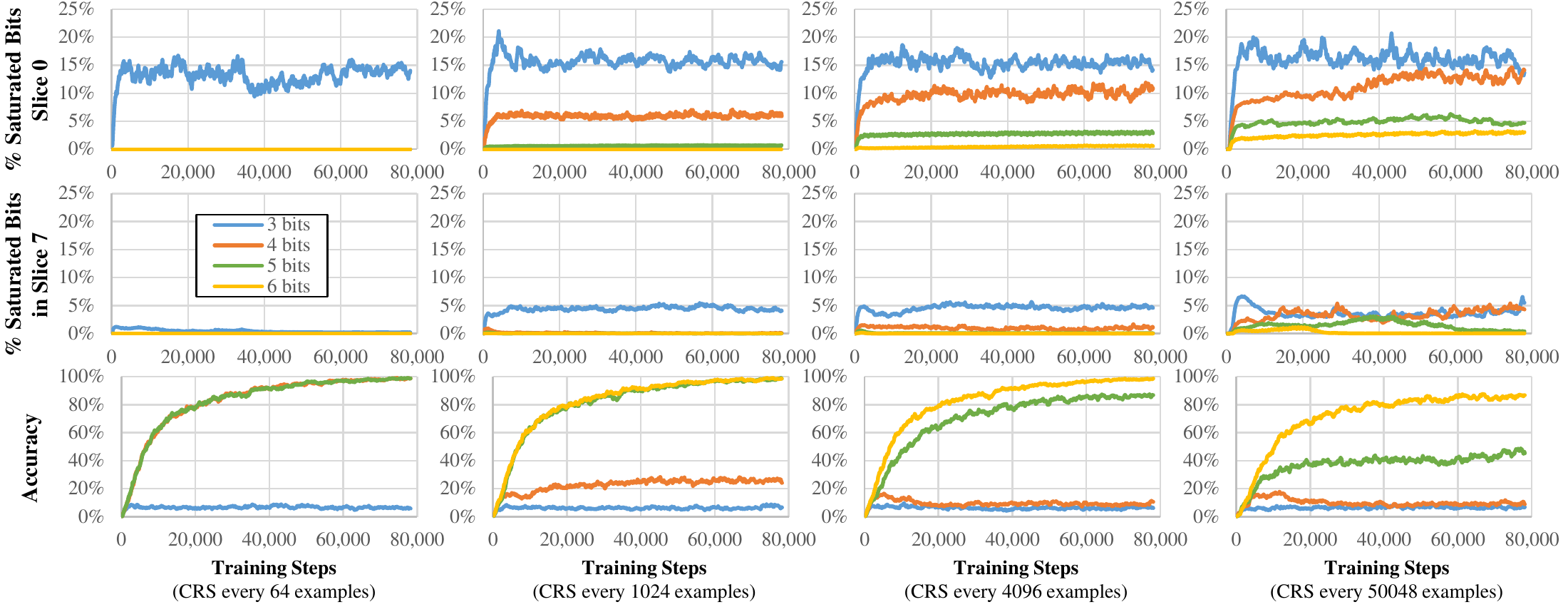}
    \vspace{-6mm}
    \caption{Impact of Slice Bits and CRS Frequency on Accuracy}\label{fig:eval-slice-crs}
    \vspace{-4mm}
\end{figure*}

\subsection {Functional Simulator}
\addition{
We implement a functional simulator using TensorFlow that models PANTHER's bit-sliced OPA technique.
This simulator enables performing design space exploration (for accuracy) on large-scale DNNs to explore the bounds on heterogeneous weight slicing and CRS frequency for trainability.
Here, a layer's weights are represented as a multi-dimensional tensor of shape $S \times M \times N$, where $S$ corresponds to a weight slice (discussed in Figure~\ref{fig:slice_op} (d)), and $M$ and $N$ correspond to the weight matrix's dimensions respectively.
Each weight slice can have a unique bit-precision, to model heterogeneous configurations (Section~\ref{sec:heterogeneous-slicing}).
The weight values beyond the range permissible by the bit-precision are clipped to model a slice's saturation.
Subsequently, the weight update operation in native TensorFlow is modified to quantize and bit-slice the computed weight gradients and then update the previous copy of weights (already quantized and bit-sliced).
Figures~\ref{fig:tensorboard-graphs} (a) and (b) show the computational graphs for an example neural network model, and the example model augmented with PANTHER OPA operation (shown in red) respectively.
}

\subsection {Baselines}

\addition{
We evaluate \reramop{} against three \additionx{weight-stationary} ASIC baselines: \cmosall{}, \rerammvm{}, and \reramopasmvm{}, as well as one NVIDIA GPU platform - Turing RTX 2080-Ti (2080-Ti).
}

\cmosall \ uses a digital version of the MCU where weights are stored in an SRAM array within the core and matrix operations are performed with a digital VFU.
\addition{
\cmosall{} is an adaptation of the digital baseline used in PUMA~\cite{ankit2019puma}.
As shown in the PUMA work, this digital baseline is an optimistic estimate of the Google TPU~\cite{jouppi2017tpu}.
It is optimistic because it uses weight-stationary MVM computations similar to TPU, but assumes that the entire model is mapped using on-chip SRAM, thereby avoiding the off-chip memory access costs in TPU.
Therefore, our comparisons with \cmosall{} also serve as a lower-bound on PANTHER's improvements compared to TPU.
}
\additionx{
The objective of comparing with \cmosall{} is to demonstrate the benefit of ReRAM-based computing over pure digital approaches.
}

\rerammvm \ uses ReRAM for MVM and \MTVM, and a digital VFU for OPA with serial reads/writes to the crossbar.
\reramopasmvm \ is a replication of PipeLayer's~\cite{song2017pipelayer} approach described in Section~\ref{sec:compare-pipelayer} and only applies to convolutional layers.
It uses ReRAM for MVM and \MTVM, and realizes OPA with ReRAM MVMs and serial reads/writes.
\additionx{
The objective of comparing with \rerammvm{} and \reramopasmvm{} is to demonstrate the benefit of ReRAM-based OPA operations.
}

\textbf{Configurations.}
\rerammvm{} and \reramopasmvm{} use 32-bit weights sliced across 16 slices with 2 bits each, which is optimal since crossbars only do MVM/\MTVM.
\reramop{} uses heterogeneous weight slicing with 32-bit weights represented using 39 bits sliced across 8 slices distributed from MSB to LSB like so: 44466555 (unless otherwise specified).
For this reason, \reramop{} consumes $17.5\%$ higher energy for MVM/\MTVM than \rerammvm{} and \reramopasmvm{} due to higher ADC precision.
We also use a CRS frequency of 1024 steps (unless otherwise specified) which achieves similar accuracy as the software implementation.
\addition{
For all three ASIC baselines and PANTHER, the hierarchical organization uses $138$ tiles per node, with $8$ cores per tile and $2$ MCUs per core.
Table~\ref{tab:hw-platforms} summarizes the platforms.
Note that both \rerammvm{} and \reramopasmvm{} have same platform parameters as PANTHER.
}

\subsection {Workloads}

\addition{
We use a 4-layered MLP model and Vgg-16 CNN model on SVHN and CIFAR-100 datasets, respectively.
Table~\ref{tab:workload} details the layer details of the two models and their computational intensity (operations to byte ratio).
The individual layers of the chosen MLP and CNN models span a wide range of computational intensity observed across the spectrum of neural network workloads.
Thus, our workloads are well representative of the large variety of layer types found in neural network models such as fully-connected, 2
D-convolution, point-wise convolution, etc.
}

Similar to other ReRAM training accelerators~\cite{cheng2017time,song2017pipelayer}, we use fixed-point arithmetic which has been shown to be successful for training large DNNs~\cite{wu2018training}.
We use the CIFAR-100 dataset for CNN which is comparable to the ImageNet dataset in terms of training difficulty~\cite{geifmanygit,bvlc-vgg-imagenet}.
However, ImageNet's large image sizes make it difficult to run the training flow without actual hardware (CIFAR-100 requires $~2$ days and ImageNet requires $~1$ month on the simulator).

%% file: fig/06-methodology/hw-platform.tex
\begin{tabular}{|l|c|c|c|}
\hline
\textbf{Parameter}       & \textbf{PANTHER (1 node)}                    & \textbf{\cmosall{} (1 node)} & \textbf{2080-Ti (1 card)}                                                 \\ \hline \hline
\textbf{SIMD lanes}      & 108 M                                        & 108 M                                                       & 4352                                      \\ \hline
\textbf{Technology} & CMOS-ReRam (32 nm)                           & CMOS (32 nm)                                                & CMOS (12 nm)                              \\ \hline
\textbf{Frequency}       & 1 GHz                                        & 1 GHz                                                       & 1.5 GHz                                   \\ \hline
\textbf{Area}            & 117 mm\textsuperscript{2} & 578 mm\textsuperscript{2}                & 750 mm\textsuperscript{2}                                                    \\ \hline
\textbf{TDP}             & 105 W                                     & 839 W                                                    & 250 W                                     \\ \hline
\textbf{On-Chip Memory}  & 72.4 MB                                    & 72.4 MB                                                   & 29.5 MB                                  \\ \hline
\end{tabular}

%% file: fig/06-methodology/workload-details.tex
\begin{tabular}{|l|c|c|c|c|c|c|c|c|}
\hline
\textbf{Layer}     & \textbf{C} & \textbf{M} & \textbf{H/W} & \textbf{R/S} & \textbf{E/F} & \textbf{Wt (MB)} & \textbf{In (MB)} & \textbf{Ops/B} \\ \hline \hline
\multicolumn{9}{|c|}{\textbf{CNN-Vgg16}}                                                                                                                          \\ \hline \hline
\textbf{Conv1}     & 3          & 64         & 32           & 3            & 32           & 0.003                 & 0.006                & 368.640                \\ \hline
\textbf{Conv2}     & 32         & 64         & 32           & 3            & 16           & 0.035                 & 0.063                & 92.160                 \\ \hline
\textbf{Conv3}     & 64         & 128        & 16           & 3            & 16           & 0.141                 & 0.031                & 209.455                \\ \hline
\textbf{Conv4}     & 128        & 128        & 16           & 3            & 8            & 0.281                 & 0.063                & 52.364                 \\ \hline
\textbf{Conv5}     & 128        & 256        & 8            & 3            & 8            & 0.563                 & 0.016                & 62.270                 \\ \hline
\textbf{Conv6}     & 256        & 256        & 8            & 3            & 8            & 1.125                 & 0.031                & 62.270                 \\ \hline
\textbf{Conv7}     & 256        & 256        & 8            & 3            & 4            & 1.125                 & 0.031                & 15.568                 \\ \hline
\textbf{Conv8}     & 256        & 512        & 4            & 3            & 4            & 2.250                 & 0.008                & 15.945                 \\ \hline
\textbf{Conv9}     & 512        & 512        & 4            & 3            & 4            & 4.500                 & 0.016                & 15.945                 \\ \hline
\textbf{Conv10}    & 512        & 512        & 4            & 3            & 2            & 4.500                 & 0.016                & 3.986                  \\ \hline
\textbf{Conv11} & 512        & 512        & 2            & 3            & 2            & 4.500                 & 0.004                & 3.997                     \\ \hline
\textbf{Conv12} & 512        & 512        & 2            & 3            & 2            & 4.500                 & 0.004                & 3.997                     \\ \hline
\textbf{Conv13}    & 512        & 512        & 2            & 3            & 1            & 4.500                 & 0.004                & 0.999                  \\ \hline
\textbf{Dense14}   & 512        & 4096       & -            & -            & -            & 4.000                 & 0.001                & 1.000                  \\ \hline
\textbf{Dense15}   & 4096       & 4096       & -            & -            & -            & 32.000                & 0.008                & 1.000                  \\ \hline
\textbf{Dense16}   & 4096       & 100        & -            & -            & -            & 0.781                 & 0.008                & 0.990                  \\ \hline \hline
\multicolumn{9}{|c|}{\textbf{MLP-L4}}                                                                                                                             \\ \hline \hline
\textbf{Dense1}    & 1024       & 256        & -            & -            & -            & 0.500                 & 0.002                & 0.996                  \\ \hline
\textbf{Dense2}    & 256        & 512        & -            & -            & -            & 0.250                 & 0.000                & 0.998                  \\ \hline
\textbf{Dense3}    & 512        & 512        & -            & -            & -            & 0.500                 & 0.001                & 0.998                  \\ \hline
\textbf{Dense4}    & 512        & 10         & -            & -            & -            & 0.010                 & 0.001                & 0.909                  \\ \hline
\end{tabular}

%% file: sec/07-evaluation.tex
\section{Evaluation}\label{sec:evaluation}

\subsection{Impact of Slice Bits and CRS Frequency on Accuracy}\label{sec:eval-bits-crs}

Figure~\ref{fig:eval-slice-crs} shows the impact of the number of bits used per slice (uniform weight slicing) and CRS frequency for the CNN benchmark.
We analyze the percentage of saturated cells per slice for a lower order and higher order slice, and their implications on CNN's Top-5 training accuracy.

Using 3 bits per slice shows significantly higher percentage of saturated cells for the lower order slice (Slice 0) than other configurations.
Further, increasing the CRS frequency does not reduce the saturation fraction of Slice 0 at 3-bits.
Consequently, the training accuracy with 3-bits slices remains very low throughout the training steps.

Using 4 bits per slice performs well at high CRS frequency (CRS every 64 steps), but does not scale well at lower CRS frequencies.
A high CRS frequency is undesirable due to the high cost of serial reads and writes incurred during carry propagation between discrete slices.

Slices with 5-bits and 6-bits are robust to repeated weight updates as they exhibit lower saturation for both lower order and higher order slices even at low CRS frequencies (every 1024 or 4096 steps).
Note that the cost of a CRS operation at low frequency (every 1024 steps) has negligible impact on overall energy and performance ($\leq4.8\%$).

Figure~\ref{fig:eval-slice-crs} also motivates heterogeneous weight slicing because it shows that the higher order slice has significantly lower saturation in general than the lower order slice.

\begin{figure}
    \centering
    \includegraphics[width=\columnwidth]{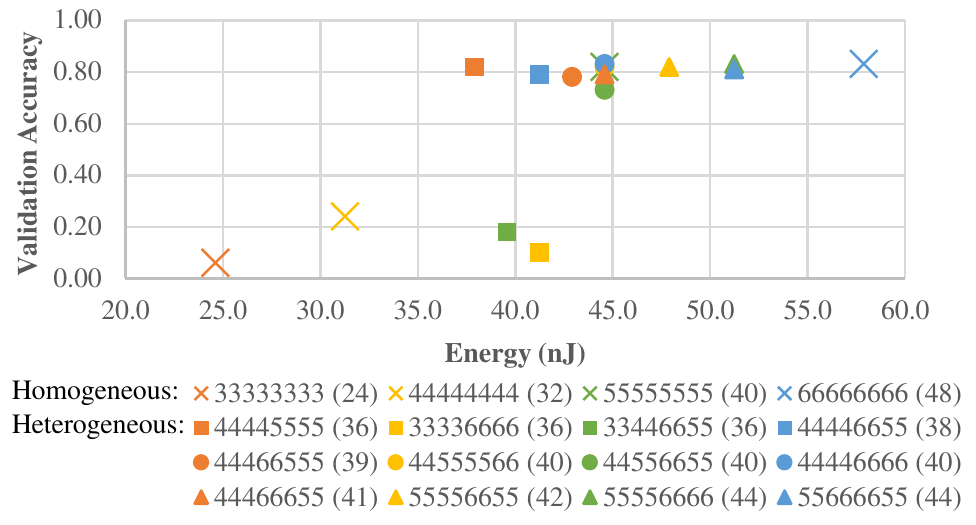}
    \vspace{-4mm}
    \caption{Heterogeneous Weight-Slicing}\label{fig:hetero-slice}
    \vspace{-4mm}
\end{figure}

\subsection{Impact of Heterogeneous Weight Slicing}\label{sec:eval-heterogeneous-slicing}

Figure~\ref{fig:hetero-slice} shows the accuracy and energy of sixteen slicing configurations.
Generally speaking, increasing the total number of bits improves accuracy by reducing saturation, but it also increases energy because it requires higher precision ADCs for MVM and \MTVM.
The graph shows that heterogeneous weight slicing enables favourable accuracy-energy tradeoffs, enabling lower energy at comparable accuracy or better accuracy at comparable energy.
Provisioning $\geq4$ bits for the four higher order slices ($4-7$) and $\geq5$ bits for the four lower order slices ($0-3$) ensures desirable accuracy.
Any configuration using 3 bit slices (irrespective of total bits) leads to significant accuracy degradation.
Note that the configuration used in the rest of the evaluation (44466555) is not a Pareto-optimal one, so our energy numbers in the rest of the evaluation are underestimated.

\begin{figure*}
    \centering
    \includegraphics[width=\textwidth]{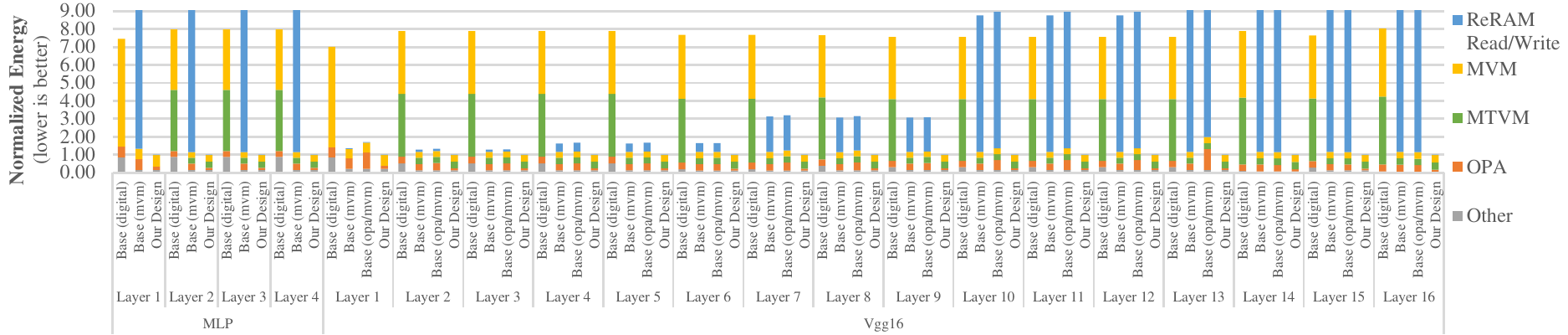}
    \vspace{-6mm}
    \caption{SGD Energy (high bars are clipped)}\label{fig:sgd-energy}
    \vspace{-2mm}
\end{figure*}

\begin{figure*}
    \centering
    \includegraphics[width=\textwidth]{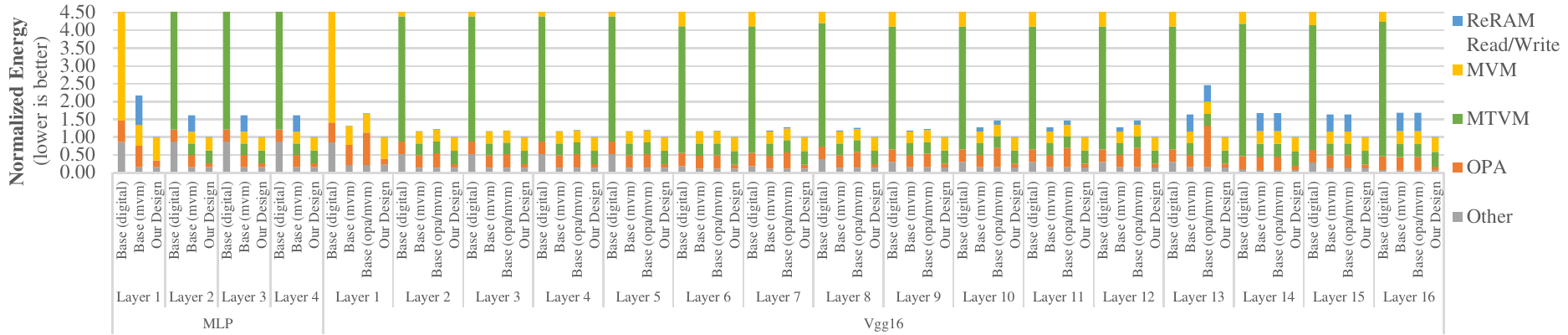}
    \vspace{-6mm}
    \caption{Mini-batch SGD Energy (high bars are clipped)}\label{fig:sgd-mb-energy}
    \vspace{-2mm}
\end{figure*}

\subsection{Variant \#1 SGD Energy Comparison} \label{sec:eval-energy-sgd}

Figure~\ref{fig:sgd-energy} compares the layer-wise energy consumption of \reramop{}'s Variant \#1  to that of all three baselines for SGD.

\textbf{\cmosall{}.} Compared to \cmosall{}, we achieve $7.01\times$--$8.02\times$ reduction in energy.
This advantage is due to the energy efficiency of computing MVM, \MTVM, and OPA in ReRAM.

\textbf{\rerammvm{}.} Compared to \rerammvm{}, we achieve $31.03\times$--$54.21\times$ reductions in energy for FC layers (Layers 1-4 in MLP and 14-16 in CNN) and $1.47\times$--$31.56\times$ for convolution layers (Layers 1-13), with the later (smaller) convolution layers showing larger reductions.
Recall that \rerammvm{} uses serial reads and writes to perform the OPA operation with digital logic.
While the large convolutional layers can amortize these reads and writes, the FC layers and small convolutional layers do not have enough work to do so which is why they suffer relatively.
In contrast, \reramop{} avoids these reads and writes by performing OPA in the crossbar ($11.37$ nJ).
\ignore{
This advantage is due to the energy efficiency of performing OPA in ReRAM compared to CMOS as well as avoiding costly ReRAM write operations to weights.
The disparity in energy reductions depends on the amount of work done per weight to amortize the ReRAM writes, where fully-connected layers have little work per weight and convolutional layers have more work per weight with the work decreasing as layers progress.
Interestingly, \cmosall{} is more energy-efficient for fully-connected layers and later convolution layers where the non-amortized cost of ReRAM writes outweighs the benefits of ReRAM MVM and \MTVM operations.
}

\textbf{\reramopasmvm{}.} \reramopasmvm{} behaves similarly to \rerammvm{}.
Recall that both baselines perform serial reads and writes to crossbars for OPA, but \rerammvm{} uses CMOS VFUs while \reramopasmvm{} uses ReRAM MVMs.
Since ReRAM MVMs and CMOS OPAs have comparable energy consumption ($35.10$ nJ and $37.28$ nJ respectively), the overall energy of the two baselines is similar.

\subsection{Variant \#2 Mini-Batch SGD Energy} \label{sec:eval-energy-smb}

Figure~\ref{fig:sgd-mb-energy} compares the layer-wise energy consumption of Variant \#2 of \reramop{} to that of all three baselines for Mini-Batch SGD with batch size 64.
Compared to SGD results (Figure~\ref{fig:sgd-energy}), the key difference is that having multiple batches before weight updates amortizes the cost of serial reads and writes in \rerammvm{} and \reramopasmvm{} (smaller blue bar).
Our energy improvements therefore come mainly from reducing OPA energy.
Energy is reduced by $1.61\times$--$2.16\times$ for fully connected layers for \rerammvm{} and \reramopasmvm{}.
It is reduced by $1.18\times$--$1.63\times$ and $1.22\times$--$2.45\times$ for convolutional layers for \rerammvm{} and \reramopasmvm{}, respectively.

For very large batch sizes such as 1,024 (not shown in the figure), ReRAM writes can be completely amortized by \rerammvm{} and \reramopasmvm{}.
In this case, \reramop{} reduces energy by $\simeq1.18\times$ compared to \rerammvm{} and \reramopasmvm{} due to reducing OPA energy.
However, batch sizes preferred by ML practitioners for DNN training ($32$, $64$) are typically smaller than what is required to amortize the ReRAM memory access costs because large batch sizes have adverse effects on DNN generalization~\cite{masters2018revisiting}.

\ignore{
\izzat{I moved this paragraph here from Section~\ref{sec:arch-mini-batch}. It is redundant with the paragraph above so I think it can be eliminated.}
Note that for accelerators that use digital logic and serial writes for weight updates, the cost of memory access to ReRAM in mini-batch SGD may be amortized over multiple input examples, as weight gradients can be computed with CMOS digital logic and written on ReRAM at the end of the batch.
However, the batch sizes desired by ML practitioners ($32$, $64$) are typically smaller than what is required to amortize the ReRAM memory access costs~\cite{masters2018revisiting}.
Thus, compared to past ReRAM accelerators, the OPA is extremely attractive for mini-batch SGD.
}

\subsection{Variant \#2 Execution Time} \label{sec:eval-perf}

\begin{figure}
    \centering
    \includegraphics[width=\columnwidth]{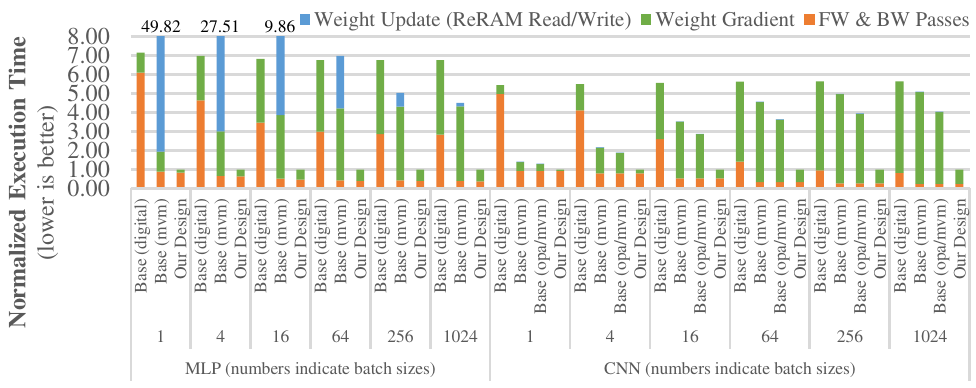}
    \caption{Execution Time}\label{fig:latency}
\end{figure}

\ignore{
\dejan{Sergey kindly agreed to read the paper and provide feedback, primarily on evaluation. His high level feedback was that latency is not as important for evaluation of training but rather throughput. That data ingestion can also be hidden by training and gotten for free if done well and that’s why latency in its own right is much less important than throughput.}
}

Figure~\ref{fig:latency} compares the layer-wise execution time of Variant \#2 to all three baselines for different batch sizes.

\textbf{\cmosall{}.} Compared to \cmosall{}, we have consistently lower execution time due to faster MVM, \MTVM, and OPA operations in ReRAM.

\textbf{\rerammvm{}.}  For MLPs with small batch sizes, \rerammvm{} significantly suffers because the ReRAM write latency is not amortized.
However, for larger batch sizes and for CNNs, the ReRAM write latency is amortized.
Nevertheless, we still outperform \rerammvm{} across all batch sizes because of lower latency ReRAM OPA.
In fact, our advantage grows with batch size because OPA consumes a larger percentage of the total time for larger batches since the forward and backward passes benefit from pipeline parallelism whereas OPA operations are serialized at the end.

\textbf{\reramopasmvm{}}. \reramopasmvm{} behaves similarly to \rerammvm{} for convolutional layers.

\begin{figure}
    \centering
    \includegraphics[width=\columnwidth]{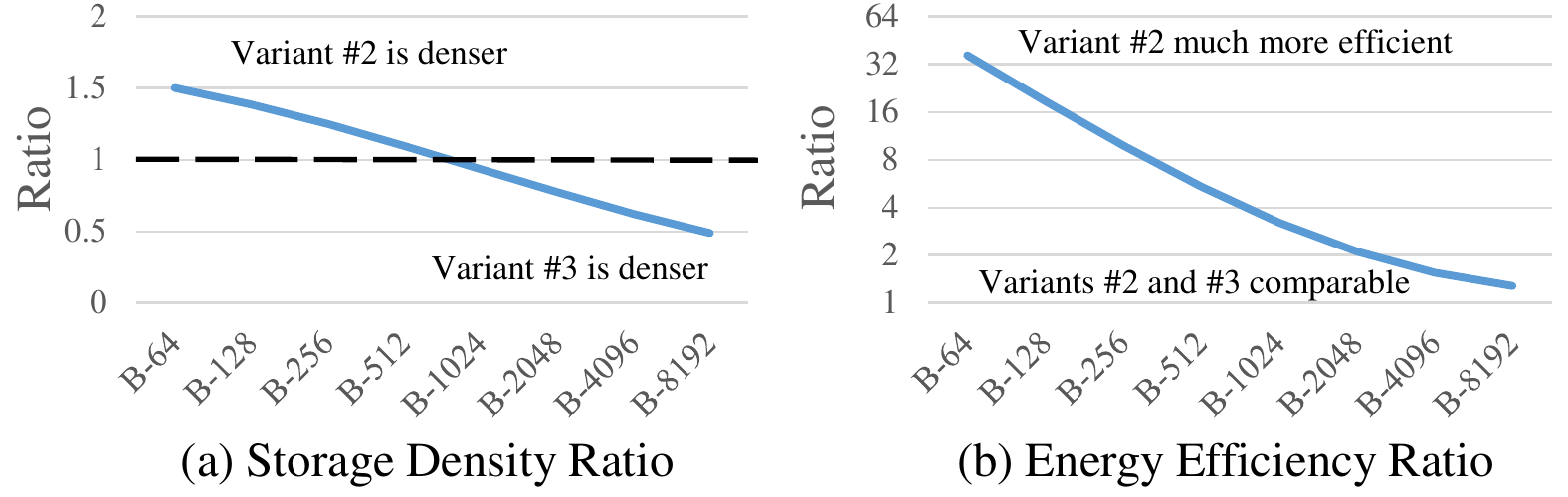}
    \vspace{-6mm}
    \caption{Variant \#2 vs. Variant \#3}\label{fig:microarch-variants-comparison}
    \vspace{-4mm}
\end{figure}

\subsection{Comparing Variants \#2 and \#3} \label{sec:panther-2x-3x-comparision}
Increasing the batch size for mini-batch SGD increases Variant \#2's shared memory requirements for storing all activations and layer gradients in the batch, degrading its storage density.
Variant \#3 uses a third crossbar for eagerly computing and storing weight gradients, thereby keeping shared memory requirements low at the expense of higher energy to commit the updates to the other crossbars at the end.
Figure~\ref{fig:microarch-variants-comparison} shows that Variant \#2 has better storage density and energy efficiency for small batch sizes, while Variant \#3 has better storage density for very large batch sizes at comparable energy efficiency.

\subsection{Comparison with GPUs}

\begin{figure}[t]
    \centering
    \includegraphics[width=\columnwidth]{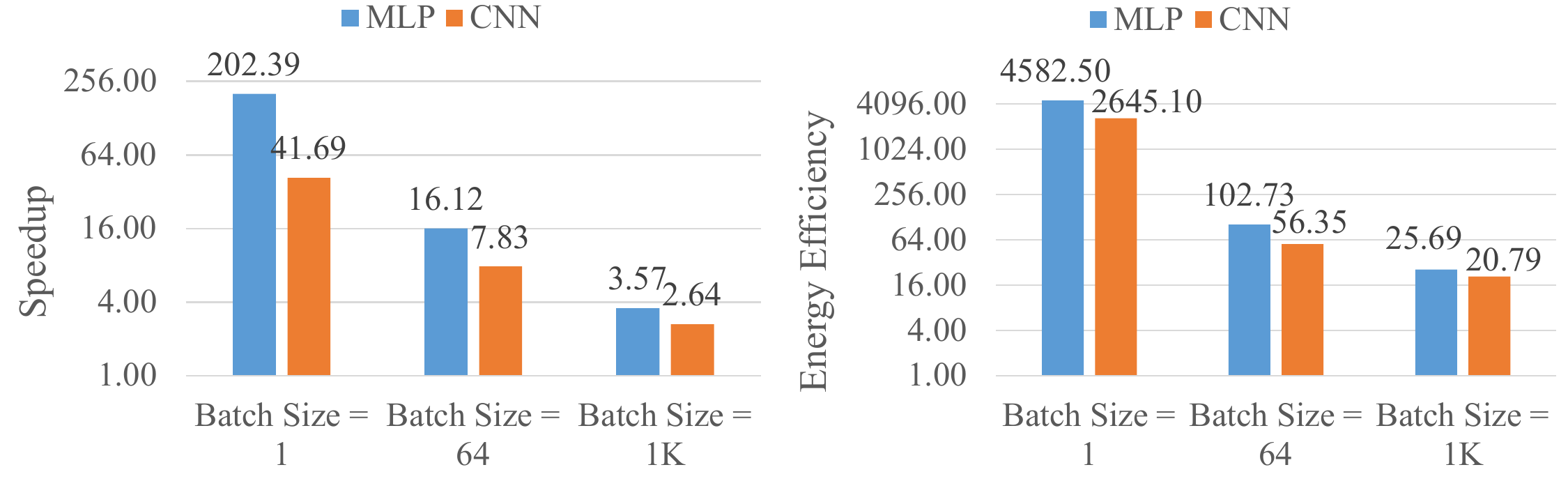}
    \vspace{-6mm}
    \caption{\addition{PANTHER's speedup and energy-efficiency compared to GPU}}\label{fig:eval-gpu}
    \vspace{-4mm}
\end{figure}

\addition{
Figure~\ref{fig:eval-gpu} compares the energy consumption and execution time of Variant \#2 with a 2080-Ti GPU for SGD (batch size 1) and Mini-Batch SGD (batch sizes 64 and 1k).
}
Our design significantly reduces energy consumption and execution time due to the use of energy-efficient and highly parallel ReRAM-based matrix operations.

GPUs rely on data reuse to hide memory access latency.
For this reason, their relative performance is worse for MLP compared compared to CNN, and for smaller batch sizes compared to larger ones.
Our design enables efficient training for a wide spectrum of batch sizes (small to large).
Training based on small batch sizes is common in emerging applications such as lifelong learning~\cite{chen2016lifelong} and online reinforcement learning~\cite{stadie2015incentivizing}, where training does not rely on any earlier collected dataset.

\subsection{Sensistivity to ReRAM endurance}
ReRAM devices have finite switching (1 to 0, 0 to 1) endurance of $10^9$ conservative writes~\cite{wei2008highly,yang2010high}, which limits their applicability towards on-chip memories for typical workloads.
However, the small magnitude of typical weight updates make ReRAM feasible for DNN training.
Considering a $5\%$ average conductance change per batch, the lifetime of a chip will be $\simeq6$ years (assuming $50\%$ reduction from failed training flows), for 1,000 trainings per year where each training is comprised of 100 epochs, 64 batch-size and 1M training examples (typical parameters in state-of-the-art image recognition benchmarks~\cite{bearpawgit}).
While weight slicing makes lower order slices more prone to degradation arising from limited endurance, adding redundancy at lower order slices and higher endurance from technology improvements (currently shown in spintronics~\cite{fong2016spin}) can make the chip more robust.

%% file: sec/08-related.tex
\section{Related Work}

Various ReRAM-based training accelerators~\cite{cheng2017time,song2017pipelayer} have been proposed, but they rely on expensive serial reads and writes to accomplish weight updates.
We avoid these reads and writes by leveraging the in-crossbar OPA operations~\cite{marinella2018multiscale,narayanan2017toward}, and extending their precision for practical trainability.
Our crossbar architecture can be used to enhance existing accelerators.

ReRAM-based accelerators have also been proposed for DNN inference~\cite{ankit2019puma,chi2016prime,liu2015reno,shafiee2016isaac}, graph processing~\cite{song2018graphr}, scientific computing~\cite{feinberg2018enabling}, and general purpose data parallel applications~\cite{fujiki2018memory}.
Our work focuses on DNN training.

Analog~\cite{likamwa2016redeye,srivastava2018promise} and DRAM-based~\cite{gao2017tetris,kim2016neurocube,li2017drisa} accelerators have been proposed as alternatives to digital-CMOS accelerators.
Our work uses ReRAM as an alternative.


Many accelerators use digital CMOS technology for accelerating DNNs, including those that mainly target inference~\cite{sze2017efficient} or also target training~\cite{venkataramani2017scaledeep}.
Our work uses hybrid digital-analog computation based on ReRAM crossbars, not just CMOS.

\addition{
Recent works have explored training DNNs with reduced precisions in floating-point arithmetic domain such as bfloat16~\cite{kalamkar2019study}, float8~\cite{wang2018training} as well as fixed-point arithmetic domain~\cite{wu2018training, yang2019training}.
While floating-point arithmetic is not amenable to ReRam-based hardware (without modifications), the reductions in fixed-point precision can be exploited in PANTHER by reducing the MCU width (number of slices) to improve training energy and time.
}

ReRAM technology suffers from imprecise writes due to non-idealities (noise and non-linearity) and manufacturability issues (stuck-at-faults and process variations).
However, the iterative nature of DNN training and careful re-training helps recover the accuracy loss from non-idealities~\cite{agarwal2016resistive}, faults~\cite{liu2017rescuing}, and variations~\cite{Chen_2017}.
Re-training is a fine-tuning process (typically 1 epoch) with insignificant cost compared to training.

%% file: sec/09-conclusion.tex
\section{Conclusion}

We propose a bit-slicing technique for enhancing the precision of ReRAM-based OPA operations to achieve sufficient precision for DNN training.
We incorporate our technique into a crossbar architecture that performs high-precision MVM and OPA operations, and present three variants catered to different training algorithms: SGD, mini-batch SGD, and mini-batch SGD with large batches.
Finally, to evaluate our design on different layer types \additionx{and training algorithms}, we develop \additionx{\reramop{}}, an ISA-programmable training accelerator with compiler support.
Our evaluation shows that \additionx{\reramop{}} achieves up to $8.02\times$, $54.21\times$, and $103\times$ energy reductions as well as $7.16\times$, $4.02\times$, and  $16\times$ execution time reductions compared to digital accelerators, ReRAM-based accelerators, and GPUs, respectively.
The proposed accelerator explores the feasibility of ReRAM technology for DNN training by mitigating their serial read and write limitations, and can pave the way for efficient design of future machine learning systems.

%% file: sec/10-ack.tex
\section*{Acknowledgement}
This work was supported by \additionx{Hewlett Packard Labs, and} the Center for Brain-inspired Computing (C-BRIC), one of six centers in JUMP, a DARPA sponsored Semiconductor Research Corporation (SRC) program.
Sandia National Laboratories is a multimission laboratory managed and operated by National Technology \& Engineering Solutions of Sandia, LLC, a wholly owned subsidiary of Honeywell International Inc., for the U.S. Department of Energy's National Nuclear Security Administration under contract DE-NA0003525. 
This paper describes objective technical results and analysis. 
Any subjective views or opinions that might be expressed in the paper do not necessarily represent the views of the U.S. Department of Energy or the United States Government.

%% file: main.bbl
\begin{thebibliography}{10}

\bibitem{sze2017efficient}
Vivienne Sze, Yu-Hsin Chen, Tien-Ju Yang, and Joel Emer.
\newblock Efficient processing of deep neural networks: A tutorial and survey.
\newblock {\em arXiv preprint arXiv:1703.09039}, 2017.

\bibitem{chen2014dadiannao}
Yunji Chen, Tao Luo, Shaoli Liu, Shijin Zhang, Liqiang He, Jia Wang, Ling Li,
  Tianshi Chen, Zhiwei Xu, Ninghui Sun, et~al.
\newblock {DaDianNao}: A machine-learning supercomputer.
\newblock In {\em Proceedings of the 47th Annual IEEE/ACM International
  Symposium on Microarchitecture}, pages 609--622. IEEE Computer Society, 2014.

\bibitem{jouppi2017tpu}
Norman~P. Jouppi, Cliff Young, Nishant Patil, David Patterson, Gaurav Agrawal,
  Raminder Bajwa, Sarah Bates, Suresh Bhatia, Nan Boden, Al~Borchers, Rick
  Boyle, Pierre-luc Cantin, Clifford Chao, Chris Clark, Jeremy Coriell, Mike
  Daley, Matt Dau, Jeffrey Dean, Ben Gelb, Tara~Vazir Ghaemmaghami, Rajendra
  Gottipati, William Gulland, Robert Hagmann, C.~Richard Ho, Doug Hogberg, John
  Hu, Robert Hundt, Dan Hurt, Julian Ibarz, Aaron Jaffey, Alek Jaworski,
  Alexander Kaplan, Harshit Khaitan, Daniel Killebrew, Andy Koch, Naveen Kumar,
  Steve Lacy, James Laudon, James Law, Diemthu Le, Chris Leary, Zhuyuan Liu,
  Kyle Lucke, Alan Lundin, Gordon MacKean, Adriana Maggiore, Maire Mahony,
  Kieran Miller, Rahul Nagarajan, Ravi Narayanaswami, Ray Ni, Kathy Nix, Thomas
  Norrie, Mark Omernick, Narayana Penukonda, Andy Phelps, Jonathan Ross, Matt
  Ross, Amir Salek, Emad Samadiani, Chris Severn, Gregory Sizikov, Matthew
  Snelham, Jed Souter, Dan Steinberg, Andy Swing, Mercedes Tan, Gregory
  Thorson, Bo~Tian, Horia Toma, Erick Tuttle, Vijay Vasudevan, Richard Walter,
  Walter Wang, Eric Wilcox, and Doe~Hyun Yoon.
\newblock In-datacenter performance analysis of a tensor processing unit.
\newblock In {\em Proceedings of the 44th Annual International Symposium on
  Computer Architecture}, ISCA'17, pages 1--12, New York, NY, USA, 2017. ACM.

\bibitem{shafiee2016isaac}
Ali Shafiee, Anirban Nag, Naveen Muralimanohar, Rajeev Balasubramonian,
  John~Paul Strachan, Miao Hu, R~Stanley Williams, and Vivek Srikumar.
\newblock {ISAAC}: A convolutional neural network accelerator with in-situ
  analog arithmetic in crossbars.
\newblock In {\em Proceedings of the 43rd International Symposium on Computer
  Architecture}, ISCA'16, pages 14--26. IEEE Press, 2016.

\bibitem{chi2016prime}
Ping Chi, Shuangchen Li, Cong Xu, Tao Zhang, Jishen Zhao, Yongpan Liu, Yu~Wang,
  and Yuan Xie.
\newblock {PRIME}: A novel processing-in-memory architecture for neural network
  computation in {ReRAM-based} main memory.
\newblock In {\em Proceedings of the 43rd International Symposium on Computer
  Architecture}, ISCA'16, pages 27--39, Piscataway, NJ, USA, 2016. IEEE Press.

\bibitem{ankit2019puma}
Aayush Ankit, Izzat~El Hajj, Sai~Rahul Chalamalasetti, Geoffrey Ndu, Martin
  Foltin, R~Stanley Williams, Paolo Faraboschi, Wen-mei~W Hwu, John~Paul
  Strachan, Kaushik Roy, and Milojicic Dejan.
\newblock Puma: A programmable ultra-efficient memristor-based accelerator for
  machine learning inference.
\newblock In {\em Proceedings of the Twenty-Fourth International Conference on
  Architectural Support for Programming Languages and Operating Systems}, pages
  715--731. ACM, 2019.

\bibitem{cheng2017time}
Ming Cheng, Lixue Xia, Zhenhua Zhu, Yi~Cai, Yuan Xie, Yu~Wang, and Huazhong
  Yang.
\newblock Time: A training-in-memory architecture for memristor-based deep
  neural networks.
\newblock In {\em Proceedings of the 54th Annual Design Automation Conference
  2017}, page~26. ACM, 2017.

\bibitem{song2017pipelayer}
Linghao Song, Xuehai Qian, Hai Li, and Yiran Chen.
\newblock Pipelayer: A pipelined {ReRAM-based} accelerator for deep learning.
\newblock In {\em High Performance Computer Architecture (HPCA), 2017 IEEE
  International Symposium on}, pages 541--552. IEEE, 2017.

\bibitem{merced2016repeatable}
Emmanuelle~J Merced-Grafals, Noraica D{\'a}vila, Ning Ge, R~Stanley Williams,
  and John~Paul Strachan.
\newblock Repeatable, accurate, and high speed multi-level programming of
  memristor 1t1r arrays for power efficient analog computing applications.
\newblock {\em Nanotechnology}, 27(36):365202, 2016.

\bibitem{marinella2018multiscale}
Matthew~J Marinella, Sapan Agarwal, Alexander Hsia, Isaac Richter, Robin
  Jacobs-Gedrim, John Niroula, Steven~J Plimpton, Engin Ipek, and Conrad~D
  James.
\newblock Multiscale co-design analysis of energy, latency, area, and accuracy
  of a {ReRAM} analog neural training accelerator.
\newblock {\em IEEE Journal on Emerging and Selected Topics in Circuits and
  Systems}, 8(1):86--101, 2018.

\bibitem{narayanan2017toward}
Pritish Narayanan, Alessandro Fumarola, Lucas~L Sanches, Kohji Hosokawa,
  SC~Lewis, Robert~M Shelby, and Geoffrey~W Burr.
\newblock Toward on-chip acceleration of the backpropagation algorithm using
  nonvolatile memory.
\newblock {\em IBM Journal of Research and Development}, 61(4/5):11--1, 2017.

\bibitem{micikevicius2017mixed}
Paulius Micikevicius, Sharan Narang, Jonah Alben, Gregory Diamos, Erich Elsen,
  David Garcia, Boris Ginsburg, Michael Houston, Oleksii Kuchaev, Ganesh
  Venkatesh, et~al.
\newblock Mixed precision training.
\newblock {\em arXiv preprint arXiv:1710.03740}, 2017.

\bibitem{wu2018training}
Shuang Wu, Guoqi Li, Feng Chen, and Luping Shi.
\newblock Training and inference with integers in deep neural networks.
\newblock {\em arXiv preprint arXiv:1802.04680}, 2018.

\bibitem{rumelhart1985learning}
David~E Rumelhart, Geoffrey~E Hinton, and Ronald~J Williams.
\newblock Learning internal representations by error propagation.
\newblock Technical report, California Univ San Diego La Jolla Inst for
  Cognitive Science, 1985.

\bibitem{Hu2018MNIST}
Miao Hu, Catherine Graves, Can Li, Yunning Li, Ning Ge, Eric Montgomery,
  Noraica Davila, Hao Jiang, R.~Stanley Williams, J.~Joshua Yang, Qiangfei Xia,
  and John~Paul Strachan.
\newblock Memristor-based analog computation and neural network classification
  with a dot product engine.
\newblock {\em Advanced Materials}, 2018.

\bibitem{liu2015reno}
Xiaoxiao Liu, Mengjie Mao, Beiye Liu, Hai Li, Yiran Chen, Boxun Li, Yu~Wang,
  Hao Jiang, Mark Barnell, Qing Wu, et~al.
\newblock {RENO}: A high-efficient reconfigurable neuromorphic computing
  accelerator design.
\newblock In {\em Design Automation Conference {(DAC)}, 2015 52nd
  ACM/EDAC/IEEE}, pages 1--6. IEEE, 2015.

\bibitem{kim2018input}
Yulhwa Kim, Hyungjun Kim, Daehyun Ahn, and Jae-Joon Kim.
\newblock Input-splitting of large neural networks for power-efficient
  accelerator with resistive crossbar memory array.
\newblock In {\em Proceedings of the International Symposium on Low Power
  Electronics and Design}, page~41. ACM, 2018.

\bibitem{truong2014new}
Son~Ngoc Truong and Kyeong-Sik Min.
\newblock New memristor-based crossbar array architecture with 50-\% area
  reduction and 48-\% power saving for matrix-vector multiplication of analog
  neuromorphic computing.
\newblock {\em Journal of semiconductor technology and science},
  14(3):356--363, 2014.

\bibitem{murmann2011adc}
Boris Murmann.
\newblock {ADC} performance survey 1997-2011.
\newblock {\em http://www. stanford. edu/\~{} murmann/adcsurvey. html}, 2011.

\bibitem{saberi2011analysis}
Mehdi Saberi, Reza Lotfi, Khalil Mafinezhad, and Wouter~A Serdijn.
\newblock Analysis of power consumption and linearity in capacitive
  digital-to-analog converters used in successive approximation {ADCs}.
\newblock {\em IEEE Transactions on Circuits and Systems I: Regular Papers},
  58(8):1736--1748, 2011.

\bibitem{nag2018newton}
Anirban Nag, Rajeev Balasubramonian, Vivek Srikumar, Ross Walker, Ali Shafiee,
  John~Paul Strachan, and Naveen Muralimanohar.
\newblock Newton: Gravitating towards the physical limits of crossbar
  acceleration.
\newblock {\em IEEE Micro}, 38(5):41--49, 2018.

\bibitem{geifmanygit}
Yonatan Geifman.
\newblock cifar-vgg.
\newblock https://github.com/geifmany/cifar-vgg/blob/master/README.md, 2018.

\bibitem{bvlc-vgg-imagenet}
BVLC.
\newblock caffe.
\newblock
  https://github.com/BVLC/caffe/wiki/Models-accuracy-on-ImageNet-2012-val,
  2017.

\bibitem{masters2018revisiting}
Dominic Masters and Carlo Luschi.
\newblock Revisiting small batch training for deep neural networks.
\newblock {\em arXiv preprint arXiv:1804.07612}, 2018.

\bibitem{chen2016lifelong}
Zhiyuan Chen and Bing Liu.
\newblock Lifelong machine learning.
\newblock {\em Synthesis Lectures on Artificial Intelligence and Machine
  Learning}, 10(3):1--145, 2016.

\bibitem{stadie2015incentivizing}
Bradly~C Stadie, Sergey Levine, and Pieter Abbeel.
\newblock Incentivizing exploration in reinforcement learning with deep
  predictive models.
\newblock {\em arXiv preprint arXiv:1507.00814}, 2015.

\bibitem{wei2008highly}
Zhiqiang Wei, Y~Kanzawa, K~Arita, Y~Katoh, K~Kawai, S~Muraoka, S~Mitani,
  S~Fujii, K~Katayama, M~Iijima, et~al.
\newblock Highly reliable taox reram and direct evidence of redox reaction
  mechanism.
\newblock In {\em 2008 IEEE International Electron Devices Meeting}, pages
  1--4. IEEE, 2008.

\bibitem{yang2010high}
J~Joshua Yang, M-X Zhang, John~Paul Strachan, Feng Miao, Matthew~D Pickett,
  Ronald~D Kelley, G~Medeiros-Ribeiro, and R~Stanley Williams.
\newblock High switching endurance in tao x memristive devices.
\newblock {\em Applied Physics Letters}, 97(23):232102, 2010.

\bibitem{bearpawgit}
Wei Yang.
\newblock pytorch-classification.
\newblock
  https://github.com/bearpaw/pytorch-classification/blob/master/TRAINING.md,
  2017.

\bibitem{fong2016spin}
Xuanyao Fong, Yusung Kim, Rangharajan Venkatesan, Sri~Harsha Choday, Anand
  Raghunathan, and Kaushik Roy.
\newblock Spin-transfer torque memories: Devices, circuits, and systems.
\newblock {\em Proceedings of the IEEE}, 104(7):1449--1488, 2016.

\bibitem{song2018graphr}
Linghao Song, Youwei Zhuo, Xuehai Qian, Hai Li, and Yiran Chen.
\newblock {GraphR}: Accelerating graph processing using {ReRAM}.
\newblock In {\em High Performance Computer Architecture (HPCA), 2018 IEEE
  International Symposium on}, pages 531--543. IEEE, 2018.

\bibitem{feinberg2018enabling}
Ben Feinberg, Uday Kumar~Reddy Vengalam, Nathan Whitehair, Shibo Wang, and
  Engin Ipek.
\newblock Enabling scientific computing on memristive accelerators.
\newblock In {\em 2018 ACM/IEEE 45th Annual International Symposium on Computer
  Architecture}, ISCA'18, pages 367--382. IEEE, 2018.

\bibitem{fujiki2018memory}
Daichi Fujiki, Scott Mahlke, and Reetuparna Das.
\newblock In-memory data parallel processor.
\newblock In {\em Proceedings of the Twenty-Third International Conference on
  Architectural Support for Programming Languages and Operating Systems}, pages
  1--14. ACM, 2018.

\bibitem{likamwa2016redeye}
Robert LiKamWa, Yunhui Hou, Julian Gao, Mia Polansky, and Lin Zhong.
\newblock {RedEye}: analog {ConvNet} image sensor architecture for continuous
  mobile vision.
\newblock In {\em Proceedings of the 43rd International Symposium on Computer
  Architecture}, ISCA'16, pages 255--266. IEEE Press, 2016.

\bibitem{srivastava2018promise}
Prakalp Srivastava, Mingu Kang, Sujan~K Gonugondla, Sungmin Lim, Jungwook Choi,
  Vikram Adve, Nam~Sung Kim, and Naresh Shanbhag.
\newblock {PROMISE}: an end-to-end design of a programmable mixed-signal
  accelerator for machine-learning algorithms.
\newblock In {\em Proceedings of the 45th Annual International Symposium on
  Computer Architecture}, ISCA'18, pages 43--56. IEEE Press, 2018.

\bibitem{gao2017tetris}
Mingyu Gao, Jing Pu, Xuan Yang, Mark Horowitz, and Christos Kozyrakis.
\newblock Tetris: Scalable and efficient neural network acceleration with 3d
  memory.
\newblock In {\em Proceedings of the Twenty-Second International Conference on
  Architectural Support for Programming Languages and Operating Systems}, pages
  751--764. ACM, 2017.

\bibitem{kim2016neurocube}
Duckhwan Kim, Jaeha Kung, Sek Chai, Sudhakar Yalamanchili, and Saibal
  Mukhopadhyay.
\newblock Neurocube: A programmable digital neuromorphic architecture with
  high-density 3d memory.
\newblock In {\em Proceedings of the ACM/IEEE 43rd Annual International
  Symposium on Computer Architecture}, ISCA'16, pages 380--392. IEEE, 2016.

\bibitem{li2017drisa}
Shuangchen Li, Dimin Niu, Krishna~T Malladi, Hongzhong Zheng, Bob Brennan, and
  Yuan Xie.
\newblock Drisa: A {DRAM}-based reconfigurable in-situ accelerator.
\newblock In {\em Proceedings of the 50th Annual IEEE/ACM International
  Symposium on Microarchitecture}, pages 288--301. ACM, 2017.

\bibitem{venkataramani2017scaledeep}
Swagath Venkataramani, Ashish Ranjan, Subarno Banerjee, Dipankar Das, Sasikanth
  Avancha, Ashok Jagannathan, Ajaya Durg, Dheemanth Nagaraj, Bharat Kaul,
  Pradeep Dubey, and Anand Raghunathan.
\newblock {ScaleDeep:} a scalable compute architecture for learning and
  evaluating deep networks.
\newblock In {\em Proceedings of the 44th Annual International Symposium on
  Computer Architecture}, ISCA'17, pages 13--26, New York, NY, USA, 2017. ACM.

\bibitem{kalamkar2019study}
Dhiraj Kalamkar, Dheevatsa Mudigere, Naveen Mellempudi, Dipankar Das, Kunal
  Banerjee, Sasikanth Avancha, Dharma~Teja Vooturi, Nataraj Jammalamadaka,
  Jianyu Huang, Hector Yuen, et~al.
\newblock A study of bfloat16 for deep learning training.
\newblock {\em arXiv preprint arXiv:1905.12322}, 2019.

\bibitem{wang2018training}
Naigang Wang, Jungwook Choi, Daniel Brand, Chia-Yu Chen, and Kailash
  Gopalakrishnan.
\newblock Training deep neural networks with 8-bit floating point numbers.
\newblock In {\em Advances in neural information processing systems}, pages
  7675--7684, 2018.

\bibitem{yang2019training}
Yukuan Yang, Shuang Wu, Lei Deng, Tianyi Yan, Yuan Xie, and Guoqi Li.
\newblock Training high-performance and large-scale deep neural networks with
  full 8-bit integers.
\newblock {\em arXiv preprint arXiv:1909.02384}, 2019.

\bibitem{agarwal2016resistive}
Sapan Agarwal, Steven~J Plimpton, David~R Hughart, Alexander~H Hsia, Isaac
  Richter, Jonathan~A Cox, Conrad~D James, and Matthew~J Marinella.
\newblock Resistive memory device requirements for a neural algorithm
  accelerator.
\newblock In {\em 2016 International Joint Conference on Neural Networks
  (IJCNN)}, pages 929--938. IEEE, 2016.

\bibitem{liu2017rescuing}
Chenchen Liu, Miao Hu, John~Paul Strachan, and Hai~Helen Li.
\newblock Rescuing memristor-based neuromorphic design with high defects.
\newblock In {\em Proceedings of the 54th Annual Design Automation Conference
  2017}, page~87. ACM, 2017.

\bibitem{Chen_2017}
Lerong Chen, Jiawen Li, Yiran Chen, Qiuping Deng, Jiyuan Shen, Xiaoyao Liang,
  and Li~Jiang.
\newblock Accelerator-friendly neural-network training: Learning variations and
  defects in {RRAM} crossbar.
\newblock In {\em Design, Automation {\&} Test in Europe Conference {\&}
  Exhibition ({DATE}), 2017}. {IEEE}, mar 2017.

\end{thebibliography}
